\documentclass[
reprint,
superscriptaddress,
amsmath,
amssymb,
aps,
pra,
floatfix,
]{revtex4-2}
\usepackage{graphicx}
\usepackage{bm}
\usepackage{hyperref}
\usepackage{xcolor}
\usepackage{xfrac}
\usepackage{physics}
\usepackage{mathrsfs}

\newcommand{\xx}[1]{}

\newcommand{\affchem}{Department of Chemistry, Purdue University, West Lafayette, Indiana 47907, USA}
\newcommand{\affphys}{Department of Physics and Astronomy, Purdue University, West Lafayette, Indiana 47907, USA}
\newcommand{\affeng}{School of Electrical and Computer Engineering, Purdue University, West Lafayette, Indiana 47907, USA}

\begin{document}
\title{Modal input-output theory for quantum nanophotonics from the first-order Maxwell operator}

\author{Ankit Kundu}
\altaffiliation{These authors contributed equally to this work.}
\affiliation{\affeng}
\author{Ishita Agarwal}
\altaffiliation{These authors contributed equally to this work.}
\affiliation{\affphys}
\author{Adhyyan S. Mansukhani}
\affiliation{\affphys}
\author{Jonathan D. Hood}
\altaffiliation{Corresponding author: hoodjd@purdue.edu}
\affiliation{\affphys}
\affiliation{\affchem}

\date{\today}

\begin{abstract}
We develop a quantum input–output theory for open photonic systems based on macroscopic quantum electrodynamics using the first-order electromagnetic Green's function. Quantum fields enter and leave the photonic system through waveguide ports, while the electromagnetic Green's function propagates the fields through the arbitrary interior of the photonic system, which may be dispersive and absorbing. The resulting relation contains a port-to-port scattering matrix and a Langevin-noise contribution from material absorption that together preserve the bosonic output commutation relations. For embedded emitters, the low-saturation and single-excitation scattering response is determined by Green's functions connecting the ports and emitters while retaining the non-Markovianity of the photonic environment. Using finite-difference time-domain simulation, we calculate the emitter-modified transmission for a nanophotonic cavity and inverse-designed coupler, providing a direct route from computed electromagnetic response to a quantum input–output model.
\end{abstract}

\maketitle

\begin{figure}[tb]
    \centering
    \includegraphics[width=1\linewidth]{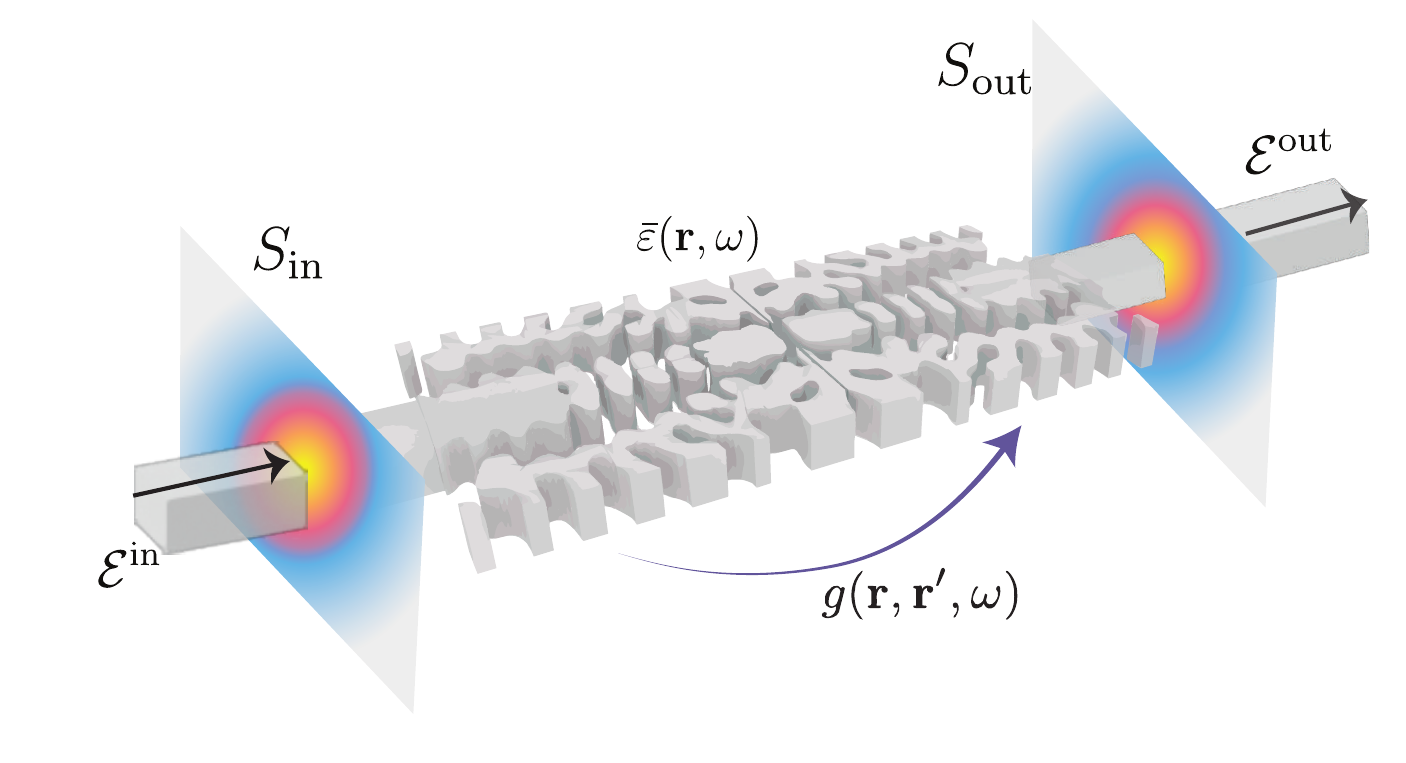}
    \caption{Illustration of the quantum input-output system for an arbitrary lossy medium characterized by $\bar{\varepsilon}(\mathbf{r},\omega)$. Input and output fields are defined on the reference planes $S_{\rm in}$ and $S_{\rm out}$. The input modes $\hat{a}_n$ are propagated into the output modes $\hat{b}_m$ by  the electromagnetic Greens function $g(\mathbf{r},\mathbf{r}',\omega)$. }
    \label{fig:illustration}
\end{figure}

\section{Introduction}
Quantum input--output theory relates light incident in specified channels to the fields leaving them, including their interaction with embedded emitters~\cite{Gardiner1985ip_op, sheremet2023waveguide}. Cavity and waveguide quantum electrodynamics (QED) commonly describe this process using selected resonances or guided modes; extending these descriptions to complex, open, and absorbing structures is less straightforward~\cite{hughes2018quantized}.  Instead, in macroscopic QED the field propagation is described by the classical electromagnetic Green's function, which can capture dispersion and absorption~\cite{gruner1996green-function,gruner1996quantum, buhmann2007dispersion}.

Macroscopic QED has been used extensively to describe Green-function-mediated interactions between emitters and their coupling to dielectric and metallic structures~\cite{novotny2006principles, martin-cano2011entanglement, asenjo-garcia2017atom, gonzalez-tudela2024light, knoll1999quantumstatetransformation}. In its conventional Langevin-noise formulation, the effects of the material absorption are encoded in noise operators. The Modified Langevin-noise formulation (MLNF) extends this by explicitly accounting for the input field operator at the boundaries, which also carries fluctuations into the system~\cite{stefano2001mode,drezet2017quantizing,ciattoni2024quantum,ciattoni2025quantum-optical, choi2025numerical, choi2026fullwavegreensfunction}. Both fluctuation channels are required to preserve the field commutation relations in finite, open systems~\cite{na2023numerical, ciattoni2026direct}. In Ref.~\cite{partI}, we generalized this formalism to finite systems whose boundaries can cut through dielectric waveguides, allowing light to enter and leave through the waveguides connected to the device.

In this paper, we develop a quantum input–output theory based on macroscopic QED where incoming and outgoing modes are defined at waveguide ports and the electromagnetic Green's function propagates the field through the interior, as shown in Fig.~\ref{fig:illustration}. This allows for an arbitrary open, dispersive, or absorptive media characterized by $\bar{\varepsilon}(\mathbf{r}, \omega) $ and does not require expanding the interior field in modes. The final result is the input-output relation
\begin{equation}
\hat b_m(\omega)=\sum_n S_{mn}(\omega)\hat a_n(\omega)+\hat\xi_m(\omega),
\end{equation}
where \(\hat a_n\) and \(\hat b_m\) annihilate photons in incoming and outgoing modes. The coefficient \(S_{mn}\) is the port-to-port scattering amplitude, while \(\hat\xi_m\) represents noise associated with material absorption.

We recover the familiar cavity- and waveguide-QED input–output relations by inserting the appropriate modal expansions of the Green's function.  We then incorporate embedded quantum emitters and derive their contribution to transmission in the low-saturation regime and the single-excitation subspace. The required electromagnetic inputs are the frequency-dependent Green's functions connecting the ports to one another, the ports to the emitters, and the emitters to one another; these can be obtained from classical solvers such as finite-difference time-domain simulations. Together, they determine the emitter-modified input–output response in this regime without constructing a separate modal model of the device.

Our method is therefore suited to complex structures such as inverse-designed couplers, photonic crystals, and metasurfaces, as well as strongly dispersive or absorbing systems such as plasmonic structures or two-dimensional materials.  Because the first-order formulation retains both electric and magnetic fields, it also treats magnetodielectric media and extends naturally to chiral media with magnetoelectric coupling.

The paper is structured as follows. We begin with a review of the first-order Green's function formalism for macroscopic QED in Sec.~\ref{subsection: Review of first-order Green-function quantization} and then show the corresponding modifications required for waveguide port boundary decompositions in Sec.~\ref{subsec: Boundary-mode-decomposition}. In Sec.~\ref{sec:applications} we show the explicit examples of our formalism for a single-mode cavity and a lossless waveguide. Then, in Sec.~\ref{sec: Coupling to emitters}, we incorporate the emitters' coupling to the nanophotonics into our input-output theory. Finally, we apply our theory using simulations for a couple of nanophotonic structures in Sec.\ref{sec:simulations} and draw our conclusions in Sec.\ref{sec: Conclusion}.

\section{Modal quantum input-output theory}

\subsection{Review of first-order Green-function quantization}
\label{subsection: Review of first-order Green-function quantization}

In our previous work~\cite{partI}, we developed a first-order Green-function quantization framework for finite photonic systems. We combine the electric and magnetic fields into the impedance-normalized dual field $\mathcal E=[\mathbf E,Z_0\mathbf H]^T$, where $Z_0=\sqrt{\mu_0/\varepsilon_0}$ is the vacuum impedance, and similarly define the dual current source $\mathcal J=[Z_0\mathbf J_E,\mathbf J_M]^T$. Here, $\mathbf J_E$ is the electric current and $\mathbf J_M$ is a fictitious magnetic current that is useful for representing equivalent surface sources. The barred notation denotes the corresponding dual-space block operators,
\begin{equation}
\bar\nabla\times=
\begin{bmatrix}
0&\nabla\times\\
-\nabla\times&0
\end{bmatrix},
\qquad
\bar\varepsilon=
\begin{bmatrix}
\boldsymbol\varepsilon&0\\
0&\boldsymbol\mu
\end{bmatrix}.
\label{eq:dual_block_operators}
\end{equation}
Defining $k_0=\omega/c$, the Hermitian operator $\mathcal H=i\bar\nabla\times$, the first-order Maxwell operator $\mathcal M=\mathcal H-k_0\bar\varepsilon$, and the dual source $\mathcal S=i\mathcal J$, Maxwell's equations take the compact form~\cite{hanson2002operator,silveirinha2024modal, ThomasWeiss2019FirstOrderPert}
\begin{equation}
\mathcal M(\omega)\mathcal E(\mathbf r,\omega)=\mathcal S(\mathbf r,\omega).
\label{eq:first_order_green_definition}
\end{equation}

The retarded Green kernel is the inverse of this first-order Maxwell operator and satisfies~\cite{silveirinha2024modal}
\begin{equation}
\mathcal M_{\mathbf r}(\omega)g(\mathbf r,\mathbf r',\omega)
=\mathbb I_6\delta^{(3)}(\mathbf r-\mathbf r')
\label{eq:first_order_green_kernel}
\end{equation}
with outgoing boundary conditions. The field generated by a dual source is therefore
\begin{equation}
\mathcal E(\mathbf r,\omega)
=\int g(\mathbf r,\mathbf r',\omega)\mathcal S(\mathbf r',\omega)\,dV'.
\label{eq:first_order_green_solution}
\end{equation}
Because $g$ propagates electric and magnetic sources into both electric and magnetic fields, it retains the directional information needed to separate incoming and outgoing channels.

As shown in Fig.~\ref{fig:illustration}, we enclose the scattering region between planar input and output reference surfaces. The medium in the interior may be dispersive and absorptive, with its electromagnetic response described by $g$. In Ref.~\cite{partI}, we showed that the quantum field operator inside this region is the sum of a volume contribution from the Langevin polarization noise $\hat{\mathcal P}_N$ associated with material absorption and a boundary contribution from the incoming field $\hat{\mathcal E}_{\rm in}$, with both contributions propagated by $g$:
\begin{multline}
\hat{\mathcal{E}}(\mathbf{r}) = k_0\!\int g(\mathbf{r},\mathbf{r}',\omega)\,\hat{\mathcal{P}}_N(\mathbf{r}')\,dV' \\
- i\oint g(\mathbf{r},\mathbf{s},\omega)\,(\bar{n}\times)\,\hat{\mathcal{E}}_\mathrm{in}(\mathbf{s})\,dS.
\label{eq:E_quantum}
\end{multline}
Treating $g$ as an integral operator, Eq.~\eqref{eq:E_quantum} can equivalently be written in bra-ket notation as
\begin{equation*}
|\hat{\mathcal E}(\omega)\rangle
=k_0g(\omega)|\hat{\mathcal P}_N(\omega)\rangle_V
-ig(\omega)(\bar n\times)|\hat{\mathcal E}_{\rm in}(\omega)\rangle_S,
\end{equation*}
where the subscripts $V$ and $S$ indicate volume and boundary integration, respectively.
The first term describes fluctuations generated by absorption inside the volume, while the second propagates the incoming boundary field into the interior. To select the field components tangential to the surface, we define the boundary-trace operator
\begin{equation}
(\bar n\times)\hat{\mathcal E}
\equiv
\begin{bmatrix}
\hat{\mathbf n}\times Z_0\hat{\mathbf H}\\
-\hat{\mathbf n}\times\hat{\mathbf E}
\end{bmatrix},
\label{eq:dual_boundary_trace}
\end{equation}
where $\hat{\mathbf n}$ is the outward surface normal. This operator removes the normal components and packages the tangential electric and magnetic fields needed to describe light entering or leaving through the boundary.

Together, the volume and boundary contributions form a self-consistent macroscopic QED description. As shown in Ref.~\cite{partI}, the generalized optical theorem combines them to give the full field commutator
\begin{multline}
[\hat{\mathcal{E}}_i(\mathbf{r},\omega),\,\hat{\mathcal{E}}_j^\dagger(\mathbf{r}',\omega')] \\
= \frac{\hbar k_0}{\pi\varepsilon_0}\,\mathrm{Im}\,g_{ij}(\mathbf{r},\mathbf{r}',\omega)\,\delta(\omega\!-\!\omega').
\label{eq:field_commutator}
\end{multline}
This is consistent with the fluctuation-dissipation theorem~\cite{callen1951irreversibility,RKubo_1966_FDT, scheel1998qed} and accounts for both bulk absorption and radiation through the boundary. The proof uses the commutators of the material polarization noise and the incoming boundary field. In the present dual-field convention, the polarization-noise commutator is~\cite{rosa2009quantumfieldsdielectriclangevin}
\begin{multline}
[\hat{\mathcal{P}}_{N,i}(\mathbf{r},\omega),\,\hat{\mathcal{P}}_{N,j}^\dagger(\mathbf{r}',\omega')]\\
= \frac{\hbar}{\pi \varepsilon_0}\,(\bar{\varepsilon}_I)_{ij}(\mathbf{r},\omega)\,\delta^{(3)}(\mathbf{r}-\mathbf{r}')\,\delta(\omega-\omega'),
\label{eq:PN_commutator}
\end{multline}
where $\bar\varepsilon_I=(\bar\varepsilon-\bar\varepsilon^\dagger)/2i$ describes material loss. The incoming boundary-field operator satisfies
\begin{multline}
\big[(\bar{n}\times\hat{\mathcal{E}}_\mathrm{in}(\mathbf{s},\omega))_i,\,(\bar{n}\times\hat{\mathcal{E}}_\mathrm{in}^\dagger(\mathbf{s}',\omega'))_j\big]\\
= -\frac{\hbar k_0}{2\pi\varepsilon_0}(\bar{n}\times)_{ij}\,\delta^{(2)}(\mathbf{s}-\mathbf{s}')\,\delta(\omega-\omega'),
\label{eq:Ein_commutator}
\end{multline}

\subsection{Boundary-mode decomposition}
\label{subsec: Boundary-mode-decomposition}

Now we extend to a modal decomposition of the fields on the boundary to create the modal input-output quantum theory. 
As shown in Appendix C of \cite{partI}, the first-order propagation relation can also be used between unbounded transverse surfaces, where the complete tangential electric and magnetic fields provide the Cauchy data for propagation through a source-free region. We therefore take the input and output reference surfaces to be infinite planes and assume that, in a small neighborhood of each plane, the material is lossless and translationally invariant along the normal direction $z$. The scattering region between the planes may still contain arbitrary structured, dispersive, and lossy media. The fields near each reference plane then have the waveguide form
\begin{equation}
\mathbf{E}(\mathbf{r}) = \mathbf{E}(\mathbf{r}_\perp)\,e^{i\beta z}, \qquad \mathbf{H}(\mathbf{r}) = \mathbf{H}(\mathbf{r}_\perp)\,e^{i\beta z},
\label{eq:modal_ansatz}
\end{equation}

Following the standard waveguide construction~\cite{Synder&love,collin1990field,svendsen2013reciprocity}, the modes $\{\mathcal E_n\}$ on each reference plane form a complete flux-orthogonal basis. Their natural overlap measures the signed electromagnetic power crossing the plane,
\begin{equation}
\begin{aligned}
\langle\mathcal{E}_m|\bar{\mathbf{z}}\times\mathcal{E}_n\rangle_S
&\equiv \int_S \mathcal{E}_m^\dagger\,(\bar{\mathbf{z}}\times)\,\mathcal{E}_n\,dS\\
&=s_m\,\delta_{mn}; \qquad s_m=\pm1.
\end{aligned}
\label{eq:flux_normalization}
\end{equation}
For a single mode, this overlap equals $-4Z_0\int_S\hat{\mathbf z}\cdot\mathbf S_{{\rm avg},m}\,dS$, where $\mathbf S_{{\rm avg},m}=\tfrac12\operatorname{Re}(\mathbf E_m\times\mathbf H_m^*)$ is the time-averaged Poynting vector~\cite{Jackson, Poynting}. With the dual-operator convention used here, $s_m=-1$ and $s_m=+1$ identify modes carrying energy in the forward and backward directions, respectively. As for an ordinary orthogonal expansion, completeness allows any boundary field to be written as

\begin{equation}
\mathcal{E}(\mathbf{r}_\perp) = \sum_n a_n\,\mathcal{E}_n(\mathbf{r}_\perp), \qquad a_n = s_n\,\langle\mathcal{E}_n|\bar{\mathbf{z}}\times\mathcal{E}\rangle_S.
\label{eq:field_expansion}
\end{equation}
In bra-ket notation, the same modal expansion is
\begin{equation*}
|\mathcal E\rangle
=\sum_n s_n|\mathcal E_n\rangle
\langle\mathcal E_n|(\bar{\mathbf z}\times)|\mathcal E\rangle_S.
\end{equation*}
The derivation of the modal eigenproblem and flux inner product is given in Appendix~\ref{app:dual_modes}.
The completeness of the channel basis also gives the identity~\cite{svendsen2013reciprocity}
\begin{equation}
\sum_n s_n\,\mathcal{E}_n(\mathbf{r}_\perp)\,\mathcal{E}_n^\dagger(\mathbf{r}'_\perp)\,(\bar{\mathbf{z}}\times) = \Pi_t\,\delta^{(2)}(\mathbf{r}_\perp - \mathbf{r}'_\perp),
\label{eq:wg_closure}
\end{equation}
where $\Pi_t$ projects onto the tangential dual field. The corresponding bra-ket resolution of the tangential identity is
\begin{equation*}
\sum_n s_n|\mathcal E_n\rangle
\langle\mathcal E_n|(\bar{\mathbf z}\times)=\Pi_t.
\end{equation*}

We now substitute the boundary-mode expansion in Eq.~\eqref{eq:field_expansion} into the quantum field relation in Eq.~\eqref{eq:E_quantum}. With $\alpha=\sqrt{2\pi\varepsilon_0/(\hbar k_0)}$, the input boundary field and output modal amplitude are normalized as $\hat{\mathcal E}_{\rm in}=\alpha^{-1}\sum_n\mathcal E_n^{\rm in}\hat a_n$ and $\hat b_m=\alpha\langle\mathcal E_m^{\rm out}|(\bar n_{\rm out}\times)|\hat{\mathcal E}\rangle_{S_{\rm out}}$. The expansion coefficients $\hat a_n(\omega)$ and $\hat b_m(\omega)$ are then quantum channel operators. We take the input operators to obey the standard bosonic commutation relations~\cite{Gardiner1985ip_op}
\begin{equation}
[\hat{a}_n(\omega),\,\hat{a}_{n'}^\dagger(\omega')] = \delta_{nn'}\,\delta(\omega - \omega').
\label{eq:a_comm}
\end{equation}

Propagating each input mode through Eq.~\eqref{eq:E_quantum} and projecting the resulting field onto output mode $m$ gives the modal input-output relation
\begin{equation}
\hat{b}_m(\omega) = \sum_n S_{mn}(\omega)\,\hat{a}_n(\omega) + \hat{\xi}_m(\omega).
\label{eq:quantum_io}
\end{equation}

\noindent where the scattering matrix $S_{mn}$ is obtained by propagating input mode $n$ with the Green's function and projecting onto output mode $m$,
\begin{equation}
\begin{aligned}
S_{mn}(\omega)
={}&-i\int_{S_{\rm out}}\!dS
\int_{S_{\rm in}}\!dS'\,
\bigl[\mathcal E_m^{\rm out}(\mathbf r,\omega)\bigr]^\dagger
(\bar n_{\rm out}\times)\\
&\quad\times g(\mathbf r,\mathbf r',\omega)
(\bar n_{\rm in}\times)
\mathcal E_n^{\rm in}(\mathbf r',\omega).
\end{aligned}
\label{eq:S-matrix}
\end{equation}
Equivalently, in bra-ket notation,
\begin{equation*}
S_{mn}(\omega)
=-i\left\langle
\mathcal E_m^{\rm out}
\middle|
(\bar n_{\rm out}\times)g(\omega)(\bar n_{\rm in}\times)
\middle|
\mathcal E_n^{\rm in}
\right\rangle_{S_{\rm out},S_{\rm in}}.
\end{equation*}
The Langevin noise operator associated with absorption inside the structure is
\begin{multline}
\hat{\xi}_m(\omega) = \alpha k_0\!\int_{S_{\rm out}}\!\!\int_V (\mathcal{E}_m^{\rm out})^\dagger\,(\bar{n}\times)\, g(\mathbf{r},\mathbf{r}',\omega) \\
\times \hat{\mathcal{P}}_N(\mathbf{r}',\omega)\,dV'\,dS.
\label{eq:noise_operator}
\end{multline}
In bra-ket notation, this becomes
\begin{equation*}
\hat\xi_m(\omega)
=\alpha k_0
\left\langle
\mathcal E_m^{\rm out}
\middle|
(\bar n_{\rm out}\times)g(\omega)
\middle|
\hat{\mathcal P}_N(\omega)
\right\rangle_{S_{\rm out},V}.
\end{equation*}

The scattering matrix in Eq.~\eqref{eq:S-matrix} is one of the key results of this paper and has a simple physical interpretation. Reading from right to left, the tangential boundary-trace operator $(\bar n\times)$ converts the input mode into the equivalent electric and magnetic surface currents. The first-order dual Green's function propagates these currents to the output plane, where the outgoing field is projected onto mode $m$ using the surface flux inner product. Thus, $S_{mn}$ is the amplitude for an input in mode $n$ to scatter into output mode $m$, with $g$ accounting for all propagation and scattering inside the structure. Equation~\eqref{eq:noise_operator} has the same output projection: the Green's function propagates the polarization noise generated by absorption throughout the volume to the output plane, and its overlap with mode $m$ gives $\hat\xi_m$.

With the scattering and noise operators now defined, we can verify that the output operators satisfy bosonic commutation relations. As shown in Appendix~\ref{app:output_commutator_full}, the Maxwell integration-by-parts identity and boundary-field representation derived in Ref.~\cite{partI}, together with port-mode orthogonality, give
\begin{equation}
[\hat{b}_m(\omega),\,\hat{b}_{m'}^\dagger(\omega')] = \delta_{mm'}\,\delta(\omega - \omega').
\label{eq:output_comm}
\end{equation}
This holds even with material loss, because the Langevin noise supplies exactly the fluctuations that absorption removes.

\section{Recovering Standard Cavity and Waveguide QED}
\label{sec:applications}

The input-output relation in Eq.~\eqref{eq:quantum_io}, together with the modal scattering matrix in Eq.~\eqref{eq:S-matrix} and the absorption-noise operator in Eq.~\eqref{eq:noise_operator}, provides a complete quantum description of linear optical scattering under the boundary assumptions above, including absorption and dispersion in the interior. The Green's function propagates the field through the structure and retains its full frequency dependence, so the description applies to both Markovian and non-Markovian electromagnetic environments. In this section, we show how the familiar cavity-QED~\cite{kimble1998strong} and waveguide-QED~\cite{sheremet2023waveguide,shen2005coherent} input-output results follow directly from these expressions.

\begin{figure}
    \centering
    \includegraphics[width=0.8\linewidth]{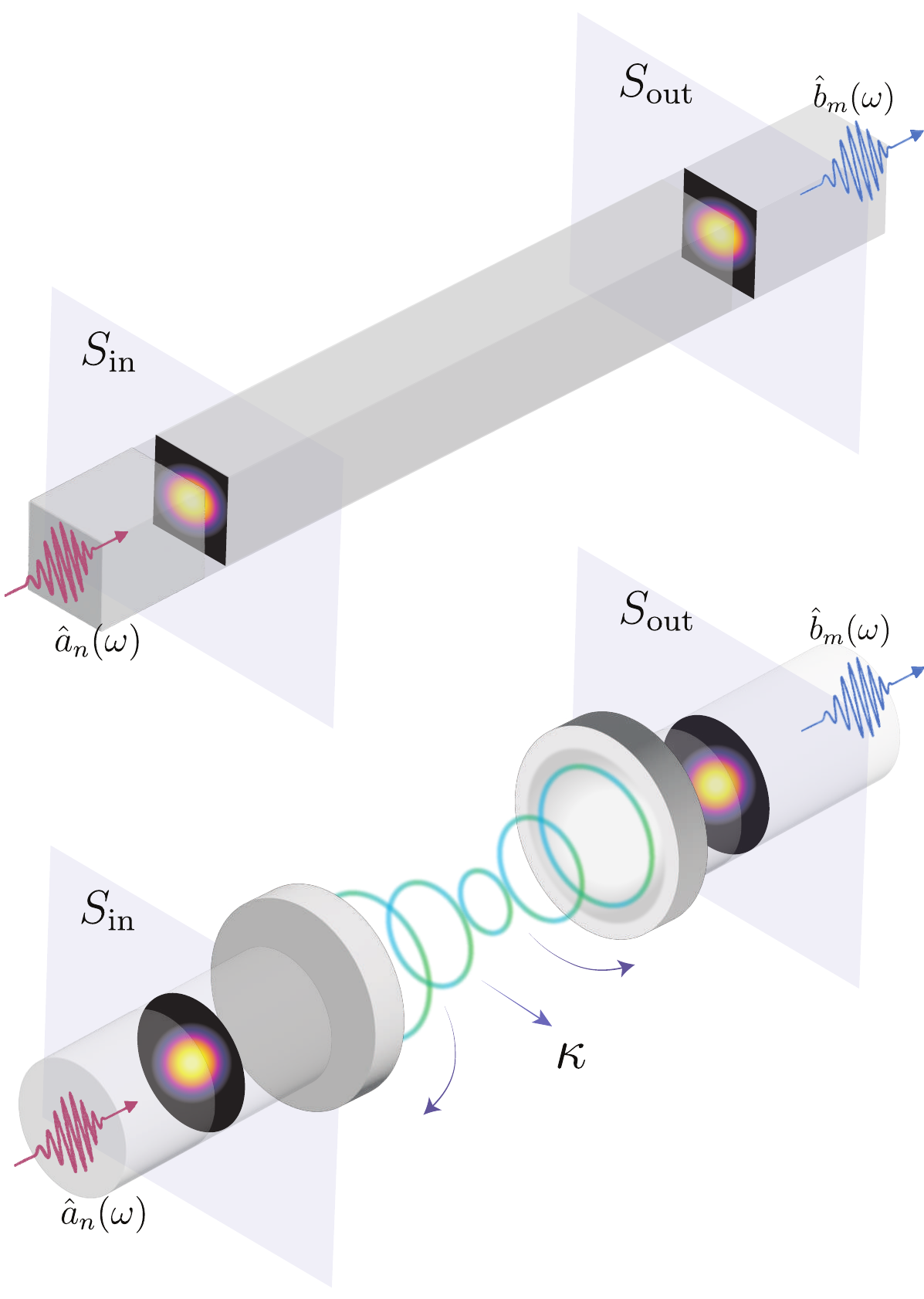}
    \caption{The two canonical geometries considered in Sec.~\ref{sec:applications}. Upper: a uniform lossless waveguide. Lower: a cavity coupled to input and output channels, with additional radiative leakage and material absorption. In both cases, the channel fields are defined on the reference planes $S_{\rm in}$ and $S_{\rm out}$.}
    \label{fig:cavity_waveguide}
\end{figure}
For a closed, lossless, nondispersive system, the first-order Green's function can be expanded in a complete set of energy-normalized eigenmodes~\cite{sakoda2004optical, silveirinha2024modal}. This simple energy normalization is restricted to the nondispersive limit but electromagnetic energy in absorbing and dispersive media requires a more general treatment~\cite{rosa2010electromagnetic,rosa2011electromagnetic}. 

Under the consideration of the material being spatially structured with restriction of closed boundary conditions to prevent energy leakage, the mode normalization derived in Appendix~\ref{app:cavity_green_derivation}, the expansion takes the form
\begin{equation}
\begin{aligned}
g(\mathbf{r},\mathbf{r}',\omega)
&=
\sum_n
\frac{
\mathcal{E}_n(\mathbf{r})\,
\mathcal{E}_n^\dagger(\mathbf{r}')
}{
k_{0,n} - k_0 - i0^+
}.
\end{aligned}
\label{eq:cav_green_spectral}
\end{equation}

For weak loss and well-separated resonances, the lossless mode profiles may be retained to leading order while the resonance frequencies acquire finite linewidths. Near a single dominant cavity resonance, as illustrated in the lower panel of Fig.~\ref{fig:cavity_waveguide}, we use the single-mode form~\cite{kristensen2014quasinormal}
\begin{equation}
g(\mathbf{r},\mathbf{r}',\omega) \approx -\frac{c\,\mathcal{E}_c(\mathbf{r})\,\mathcal{E}_c^\dagger(\mathbf{r}')}{\Delta(\omega) + i\kappa/2},
\label{eq:cav_single_mode}
\end{equation}

\noindent where $\Delta(\omega)=\omega-\omega_c$ and $\kappa$ includes both radiative leakage and material absorption. Substituting Eq.~\eqref{eq:cav_single_mode} into the modal scattering expression, Eq.~\eqref{eq:S-matrix}, separates the response into an overlap that couples input mode $n$ into the cavity and an overlap that couples the cavity to output mode $m$. We define these channel-coupling amplitudes as
\begin{equation}
\begin{aligned}
\eta_n (\omega) 
&=\sqrt{c}\,\langle\mathcal E_c| (\bar n_{\rm in}\times)|\mathcal E_n^{\rm in}\rangle_{S_{\rm in}},\\
\eta_m (\omega) 
&=i\sqrt{c}\,\langle\mathcal E_m^{\rm out}|(\bar n_{\rm out}\times)|\mathcal E_c\rangle_{S_{\rm out}}.
\end{aligned}
\label{eq:cavity_channel_couplings}
\end{equation}
The resonant contribution to the scattering matrix is therefore
\begin{equation}
S_{mn}^{\rm cav}(\omega)
=\frac{\eta_m (\omega) \eta_n (\omega) }{\omega-\omega_c+i\kappa/2}.
\label{eq:cavity_scattering_overlap}
\end{equation}
For transmission between two distinct ports of the cavity, these coupling amplitudes can be parameterized in terms of the corresponding partial cavity decay rates as
$\eta_n (\omega)=\sqrt{\kappa_{\rm in}}\,e^{i\theta_n},  \,\eta_m (\omega) =\sqrt{\kappa_{\rm out}}\,e^{i\theta_m}$,
for energy-normalized cavity and flux-normalized port modes. Combining the coupling phases into a single phase $\phi=\theta_m +\theta_n$, Eq.~\eqref{eq:cavity_scattering_overlap} reduces to
\begin{equation}
t_{\rm cav}(\omega) = \frac{e^{i\phi}\sqrt{\kappa_\mathrm{in}\kappa_\mathrm{out}}}{\omega-\omega_c + i\kappa/2}.
\label{eq:cavity_io}
\end{equation}
Here $\kappa$ accounts for leakage into all external channels and material absorption. This recovers the familiar cavity transmission response directly from the Green-function scattering expression~\cite{Gardiner1985ip_op,kimble1998strong,fan2003temporal, scully1997quantum, Jon_thesis}.

We next consider the uniform, lossless waveguide shown in the upper panel of Fig.~\ref{fig:cavity_waveguide}. Each propagating mode retains its transverse profile and accumulates a phase $e^{i\beta_n(\omega)(z-z')}$, where $\beta_n$ is its propagation constant. The guided-mode contribution to the first-order Green's function is a sum over these modes, with a step function selecting propagation away from the source in the direction of each mode's energy flow~\cite{collin1990field,felsen1994radiation, Arfken}:
\begin{multline}
g_\mathrm{guided}(\mathbf{r},\mathbf{r}',\omega) = i\sum_{n\in{\rm guided}}\Theta\!\bigl(\tau_n(z\!-\!z')\bigr) \\
\times \mathcal{E}_n(\mathbf{r}_\perp)\,\mathcal{E}_n^\dagger(\mathbf{r}'_\perp)\,e^{i\beta_n(z-z')}.
\label{eq:guided_green}
\end{multline}

\noindent Here $\tau_n=-s_n$ is the direction of energy flow. Appendix~\ref{app:waveguide_green} derives this expression from the transverse resolvent and the retarded pole prescription. Projecting the propagated field onto the output modes and using flux orthogonality gives the conventional propagation law, with the physical outgoing-amplitude convention specified in the appendix,
\begin{equation}
S_{mn}(\omega) = e^{i\beta_n(\omega)L}\,\delta_{mn},
\label{eq:S_waveguide}
\end{equation}

\noindent where $L$ is the distance between the reference planes. Thus, each mode acquires only its propagation phase, with no crosstalk between distinct modes. Since the waveguide is lossless, the absorption-noise term vanishes, and the quantum input-output relation reduces to
\begin{equation}
\hat{b}_m(\omega) = e^{i\beta_m(\omega)L}\,\hat{a}_m(\omega),
\label{eq:wg_io}
\end{equation}
recovering the usual propagation relation of waveguide QED in the absence of loss and scattering pathways~\cite{shen2005coherent,sheremet2023waveguide}. 

Appendix~\ref{app:composite_moved} applies the same construction to two coupled cavities and coupled cavity-waveguide systems, extending the applicability by using the quasi-normal mode approach~\cite{HughesCoupledCavityQED, Hughes2004SingleQD}. The compact coupled-mode Green's function can be inserted directly into Eq.~\eqref{eq:S-matrix} to evaluate input-output relations pertaining to hybrid nanophotonic systems when the two modes interact through direct spatial overlap~\cite{Hartmann2008QuantumManyBody}, photon tunneling~\cite{Knight1997Evanescentcoupling}, or radiative mixing~\cite{Horng2019RadiativeCoupling} via a shared environment.

\section{Coupling to emitters}
\label{sec: Coupling to emitters}

\begin{figure}[h!]
    \centering
    \includegraphics[width=1\linewidth]{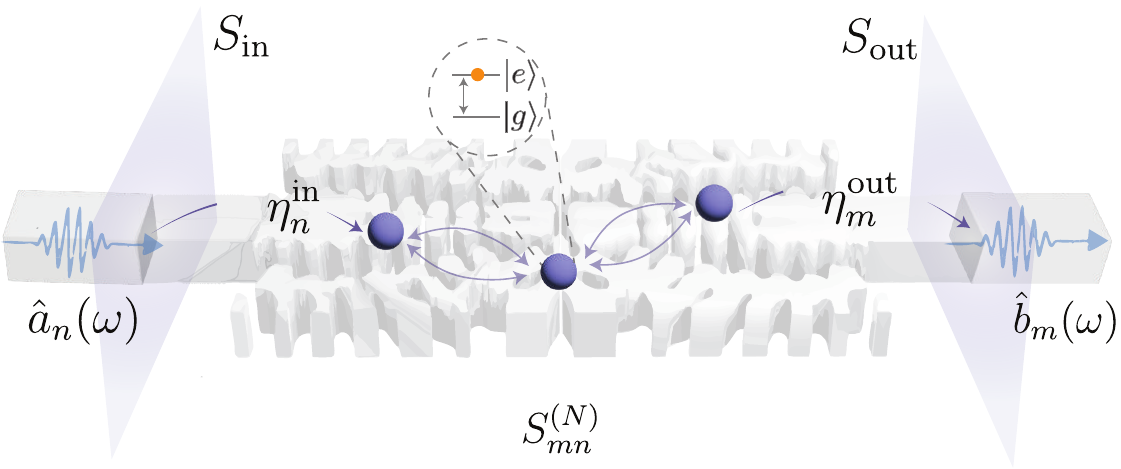}
    \caption{Two-level emitters embedded in a structured photonic environment. The input mode $\hat a_n$ drives the emitters through $\boldsymbol\eta_n^{\rm in}$, their collective response is described by $\mathbf M^{-1}$, and $\boldsymbol\eta_m^{\rm out}$ couples their radiation to output mode $\hat b_m$. This emitter-mediated contribution combines with the background scattering to give $S_{mn}^{(N)}$ in Eq.~\eqref{eq:new_dressed_S_N}.}
    \label{fig:emitter}
\end{figure}

We now extend the input-output description to include $N$ two-level emitters embedded in the photonic structure, as illustrated in Fig.~\ref{fig:emitter}. The emitters have the free Hamiltonian $\hat H_A=\sum_i\hbar\omega_i\hat\sigma_i^\dagger\hat\sigma_i$ and, in the dipole and rotating-wave approximations~\cite{Gerry_Knight_2004}, interact with the electromagnetic field through
\begin{equation}
\hat H_{\rm int}
=-\sum_i\left[
\hat\sigma_i^\dagger\mathcal D_i^\dagger\hat{\mathcal E}(\mathbf r_i,t)
+\mathrm{H.c.}
\right],
\label{eq:new_emitter_hamiltonian}
\end{equation}
where $\hat\sigma_i$ lowers emitter $i$ and $\mathcal D_i=[\mathbf d_i,\mathbf 0]^T$ contains its electric transition dipole strength. The lower block of $\mathcal D_i$ is zero because we only consider electric-dipole coupling; it would be nonzero for a magnetic transition.
The total field is the sum of the background field and the field radiated by the emitters,
\begin{equation}
\hat{\mathcal E}(\mathbf r,\omega)
=\hat{\mathcal E}_0(\mathbf r,\omega)
+\omega Z_0\sum_{j=1}^{N}g(\mathbf r,\mathbf r_j,\omega)\mathcal D_j\hat\sigma_j(\omega),
\label{eq:new_total_field_emitter}
\end{equation}
where $\hat{\mathcal E}_0$ is the field of the photonic structure void of emitters, including contributions from both the incoming field and material noise.

Applying the same output-mode projection used to obtain Eq.~\eqref{eq:quantum_io}, the background field again gives the scattering and material-noise terms, while the field radiated by the emitters gives an additional contribution to output mode $m$:
\begin{equation}
\hat b_m(\omega)
=\sum_n S_{mn}(\omega)\hat a_n(\omega)
+\hat\xi_m(\omega)
+[\boldsymbol\eta_m^{\rm out}(\omega)]^{T}\hat{\boldsymbol\sigma}(\omega).
\label{eq:exact_output_with_emitters}
\end{equation}
Here, $\hat{\boldsymbol\sigma}=(\hat\sigma_1,\ldots,\hat\sigma_N)^T$ collects the emitter lowering operators. The projection of the field emitted by emitter $j$ onto output mode $m$ is
\begin{equation}
[\boldsymbol\eta_m^{\rm out}(\omega)]_j
=\alpha\omega Z_0
\left\langle
\mathcal E_m^{\rm out}
\middle|
(\bar n_{\rm out}\times)g(\omega)
\middle|
\mathcal D_j
\right\rangle_{S_{\rm out}},
\label{eq:new_eta_out_braket}
\end{equation}
where $g(\omega)|\mathcal D_j\rangle$ denotes the field produced by the point dipole $\mathcal D_j$ at $\mathbf r_j$. 
Equations~\eqref{eq:new_total_field_emitter} and~\eqref{eq:exact_output_with_emitters} do not use either the low-saturation or Markov approximation.

The remaining problem is to determine the emitter operators $\hat{\boldsymbol\sigma}$. Their general dynamics are nonlinear and can be obtained from the Heisenberg-Langevin equations. Here, we give the solution in the low-saturation regime, where each emitter remains predominantly in its ground state and the equations become linear. The same linear equations are exact for single-photon scattering within the single-excitation subspace. Substituting the total field in Eq.~\eqref{eq:new_total_field_emitter} into the linearized emitter equations gives
\begin{equation}
\begin{aligned}
\sum_j M_{ij}(\omega)\hat\sigma_j(\omega)
&=-\frac{1}{\hbar}\mathcal D_i^\dagger
\hat{\mathcal E}_0(\mathbf r_i,\omega)\\
&=\sum_n[\boldsymbol\eta_n^{\rm in}(\omega)]_i\hat a_n(\omega)
+[\hat{\mathbf F}_N(\omega)]_i.
\end{aligned}
\label{eq:new_collective_emitter_system}
\end{equation}
which relates the emitter transition to the background field of the photonic structure and noise from the environment. Here, the collective response matrix, $\mathbf M$, and Green-function-mediated self-energy, $\Sigma$, are
\begin{equation}
\begin{aligned}
M_{ij}(\omega)
&=(\omega-\omega_i)\delta_{ij}-\Sigma_{ij}(\omega),\\
\Sigma_{ij}(\omega)
&=-\frac{\omega Z_0}{\hbar}
\mathcal D_i^\dagger g(\mathbf r_i,\mathbf r_j,\omega)\mathcal D_j.
\end{aligned}
\label{eq:new_emitter_response_matrix}
\end{equation}
The diagonal elements describe the dressed response of each emitter, while the off-diagonal elements describe interactions between different emitters mediated by the photonic environment. 

Inverting Eq.~\eqref{eq:new_collective_emitter_system} gives
\begin{equation}
\hat{\boldsymbol\sigma}(\omega)
=\mathbf M^{-1}(\omega)
\left[
\sum_n\boldsymbol\eta_n^{\rm in}(\omega)\hat a_n(\omega)
+\hat{\mathbf F}_N(\omega)
\right].
\label{eq:new_sigma_solution}
\end{equation}
The input coupling propagates input mode $n$ from the boundary to emitter $i$ and projects the resulting local field onto its transition dipole,
\begin{equation}
\begin{aligned}
[\boldsymbol\eta_n^{\rm in}(\omega)]_i
&=\frac{i}{\hbar\alpha}
\left\langle
\mathcal D_i
\middle|
g(\omega)(\bar n_{\rm in}\times)
\middle|
\mathcal E_n^{\rm in}
\right\rangle_{S_{\rm in}}\\
&=\frac{i}{\hbar\alpha}\mathcal D_i^\dagger
\int_{S_{\rm in}}g(\mathbf r_i,\mathbf s,\omega)
(\bar n_{\rm in}\times)\mathcal E_n^{\rm in}(\mathbf s,\omega)\,dS.
\end{aligned}
\label{eq:new_eta_in_definition}
\end{equation}
The material-noise drive is explicitly
\begin{equation}
\begin{aligned}
[\hat{\mathbf{F}}_N(\omega)]_i
&=-\frac{k_0}{\hbar}\mathcal D_i^\dagger
\int_V g(\mathbf r_i,\mathbf r',\omega)
\hat{\mathcal P}_N(\mathbf r',\omega)\,dV'\\
&=-\frac{k_0}{\hbar} \langle\mathcal{D}_i | g(\omega) | \hat{\mathcal{P}}_N\rangle_V.
\end{aligned}
\label{eq:new_F_N_definition}
\end{equation}

Substituting the solution for $\hat{\boldsymbol\sigma}$ from Eq.~\eqref{eq:new_sigma_solution} into the exact output relation, Eq.~\eqref{eq:exact_output_with_emitters}, and grouping the terms proportional to each input operator gives, with the common frequency argument suppressed in the next two equations,
\begin{equation}
\begin{aligned}
\hat b_m
={}&\sum_n\Bigl[S_{mn}
+(\boldsymbol\eta_m^{\rm out})^{T}\mathbf M^{-1}
\boldsymbol\eta_n^{\rm in}\Bigr]\hat a_n\\
&+\hat\xi_m
+(\boldsymbol\eta_m^{\rm out})^{T}\mathbf M^{-1}
\hat{\mathbf F}_N.
\end{aligned}
\label{eq:new_output_grouped}
\end{equation}
The coefficient of $\hat a_n$ defines the coherent scattering matrix of the photonic structure with the $N$ emitters,
\begin{equation}
S_{mn}^{(N)}
=S_{mn}
+(\boldsymbol\eta_m^{\rm out})^{T}\mathbf M^{-1}
\boldsymbol\eta_n^{\rm in}.
\label{eq:new_dressed_S_N}
\end{equation}
In Eq.~\eqref{eq:new_dressed_S_N}, $S_{mn}$ describes background scattering, while the second term describes input excitation of the emitters, their collective response through $\mathbf M^{-1}$, and radiation into output mode $m$. Equation~\eqref{eq:new_output_grouped} also retains the quantum noise: $\hat\xi_m$ is material noise propagated directly to the output, whereas $(\boldsymbol\eta_m^{\rm out})^T\mathbf M^{-1}\hat{\mathbf F}_N$ describes the same material noise driving the emitters and subsequently radiating into the output. These two contributions arise from the same material bath and are generally correlated.

Together, Eqs.~\eqref{eq:new_output_grouped} and~\eqref{eq:new_dressed_S_N} give the modal quantum input-output relation for the coupled emitter-photon system in the linear regime. The scattering matrix determines coherent transmission and reflection in the low-saturation regime, or single-photon scattering within the single-excitation subspace, while the noise terms complete the operator relation. The frequency-dependent coefficients in these relations require Green's functions connecting input modes to emitters, emitters to one another, emitters to output modes, and input modes directly to output modes; the material-noise couplings follow from the same Green's function and absorptive response.

For a single emitter and a selected input and output channel, $S_{mn}^{(N)}$ reduces to the complex transmission amplitude $t_A$. The frequency-dependent self-energy is
\begin{equation}
\Sigma_A(\omega)
=-\frac{\omega Z_0}{\hbar}\mathcal D_A^\dagger g(\mathbf r_A,\mathbf r_A,\omega)\mathcal D_A
\equiv\Delta_A(\omega)-\frac{i}{2}\Gamma_A(\omega).
\label{eq:new_single_emitter_self_energy}
\end{equation}
Here, $\Delta_A=\operatorname{Re}\Sigma_A$ is the frequency shift and $\Gamma_A=-2\operatorname{Im}\Sigma_A$ is the total electromagnetic decay rate.

Equation~\eqref{eq:new_dressed_S_N} then gives the non-Markovian transmission amplitude~\cite{asenjo-garcia2017atom}
\begin{equation}
t_A(\omega)
=t(\omega)
+\frac{\eta_{\rm out}(\omega)\eta_{\rm in}(\omega)}
{\omega-\omega_A-\Delta_A(\omega)+i\Gamma_A(\omega)/2},
\label{eq:new_single_emitter_transmission}
\end{equation}
where $t(\omega)=S_{mn}(\omega)$ is the transmission amplitude of the photonic structure without the emitter. The Markov approximation is recovered when the self-energy varies slowly across the emitter linewidth, so that $\Delta_A(\omega)$ and $\Gamma_A(\omega)$ may be evaluated at a reference frequency near resonance and treated as constants.

For $N$ emitters, the scattering can also be resolved into collective emitter modes~\cite{asenjo-garcia2017atom}. When $\mathbf M(\omega)$ is diagonalizable, its left and right eigenvectors form a biorthogonal basis. Projecting $\boldsymbol\eta_n^{\rm in}$ onto the left eigenvectors gives the excitation of each collective mode, while projecting the right eigenvectors onto $(\boldsymbol\eta_m^{\rm out})^T$ gives its radiation into the selected output. Each mode contributes the product of these overlaps divided by its eigenvalue of $\mathbf M$, so the observed transmission depends both on the collective response and on how each mode couples to the input and output channels. This decomposition may be performed at each frequency, retaining the frequency dependence of the photonic environment.

\section{Simulations}
\label{sec:simulations}

\begin{figure*}[tb]
    \centering    \includegraphics[width=\textwidth]{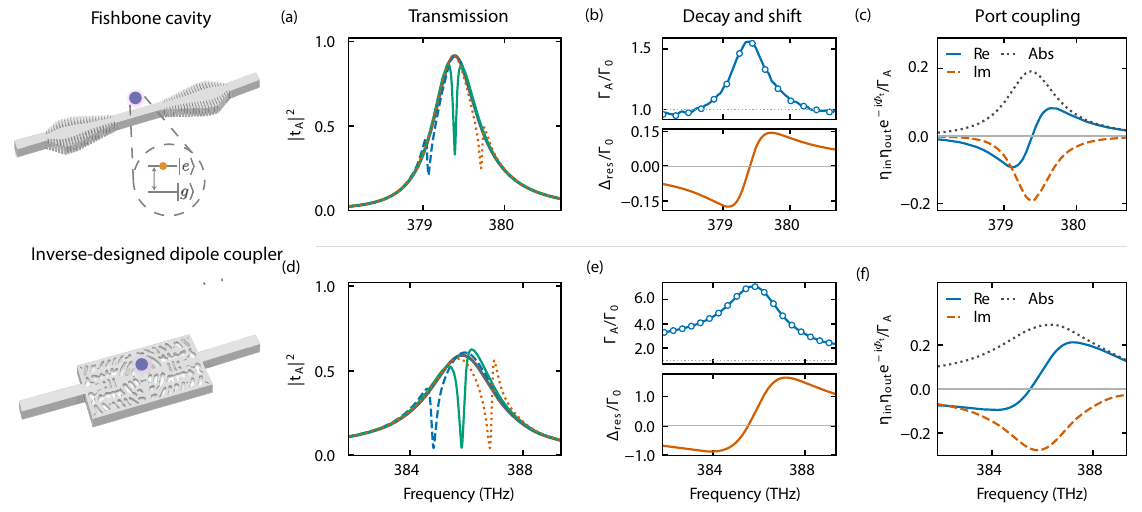}
    \caption{Single-emitter response reconstructed from classical Green-function simulations for a fishbone cavity (top row) and a bidirectional DC structure (bottom row). First column: illustrative device schematics; the multi-emitter drawing is a layout placeholder, while all curves describe one emitter. Second column: power transmission with no emitter (gray) and for three emitter detunings $\delta f_A=f_A-f_c$, with fixed free-space linewidth $\Gamma_0/(2\pi)=32.63$ GHz. The reference frequencies are $f_c=379.3918$ and $385.8333$ THz, respectively. (b,e) Upper: total electromagnetic decay rate; circles indicate every fourth original frequency sample. Lower: resonant frequency shift obtained by fitting the complex, reference-subtracted self Green's function to a retarded pole plus a smooth linear background and retaining only the pole contribution. The same shift convention is used in the transmission curves; the absolute dielectric Lamb shift is not included. (c,f) Real part (solid), imaginary part (dashed), and magnitude (dotted) of $\eta_{\rm in}\eta_{\rm out}e^{-i\phi_t}/\Gamma_A$, with $\phi_t=\arg t$, referenced to the phase of the bare transmission $t$. The response is reconstructed assuming single-mode port coupling and common source normalization; narrow emitter features are predictions from interpolated Green's functions, not directly resolved emitter simulations.}
    \label{fig:device_comparison}
\end{figure*}

Equations~\eqref{eq:new_output_grouped} and~\eqref{eq:new_dressed_S_N} provide, respectively, the quantum input-output relation and the coherent scattering amplitudes in the low-saturation regime or single-excitation subspace. We now show how Green's functions obtained from finite-difference time-domain (FDTD) simulations can be used to predict transmission in the presence of emitters. Equation~\eqref{eq:new_output_grouped} also retains both contributions from material absorption noise required by the fluctuation-dissipation relation. Applying this description to strongly dispersive and lossy systems, such as plasmonic cavities, is an interesting direction for future work. Here we illustrate the numerical construction for lossless dielectric structures: a fishbone cavity with air-cladding, and an inverse-designed dipole coupler.

We use the Tidy3D FDTD solver~\cite{tidy3d} for the simulations. The electric fields generated by dipole sources give the second-order electric Green tensor $g^{(2)}$ after source normalization. The first-order electric field tensor due to an electric dipole source follows the relation~\cite{partI}
\begin{equation}
g_{EE}(\mathbf r,\mathbf r',\omega)
=k_0g^{(2)}(\mathbf r,\mathbf r',\omega),
\label{eq:simulation_green_conversion}
\end{equation}
where $k_0=\omega/c$ and $g_{EE}$ is the $3\times3$ electric-electric block of the first-order tensor. Since $\mathcal D_i=(\mathbf d_i,\mathbf 0)^T$ has no magnetic part, electric-dipole coupling selects only this block:
\begin{equation}
\begin{aligned}
&\mathcal D_i^\dagger g(\mathbf r_i,\mathbf r_j,\omega)\mathcal D_j
=k_0\mathbf d_i^\dagger
g^{(2)}(\mathbf r_i,\mathbf r_j,\omega)\mathbf d_j.
\end{aligned}
\label{eq:simulation_dipole_conversion}
\end{equation}
Thus the second-order electric Green tensor obtained from the simulations directly determines the emitter self-energy in Eq.~\eqref{eq:new_emitter_response_matrix}.

    For a single emitter, the transmission is determined by the bare modal transmission amplitude, $t(\omega)$, the emitter self-response, $g(\mathbf{r}_o, \mathbf{r}_o, \omega)$, and propagation between the emitter and the input(output) channels, $\eta_{\mathrm{in(out)}}$. We obtain these quantities from dipole-source simulations and a separate calculation of the bare modal transmission. Although $\eta_{\rm in}$ and $\eta_{\rm out}$ involve surface-mode integrals, their product can be reconstructed from point-to-point Green's functions when a single guided mode dominates at each reference position: the input-launch and output-sampling factors cancel in the ratio $g_{\rm out}^{(2)}g_{\rm in}^{(2)}/g_{\rm in-out}^{(2)}$. We specify the emitter strength through its free-space decay rate $\Gamma_0$, while the simulated self-response determines its environment-dependent decay and frequency shift. Appendix~\ref{app:numerical_reconstruction} gives the reconstruction and normalization procedure.

Figure~\ref{fig:device_comparison} shows the reconstructed response for the fishbone cavity and the inverse-designed bidirectional dipole emitter coupler. For each structure, we compare the bare transmission with the emitter-modified transmission at three transition frequencies, keeping the free-space linewidth $\Gamma_0$ fixed. We also show the total decay enhancement $\Gamma_A/\Gamma_0$, the resonant frequency shift $\Delta_{\rm res}/\Gamma_0$, and the phase-referenced coupling $Q=\eta_{\rm out}\eta_{\rm in}e^{-i\phi_t(\omega)}/\Gamma_A$. Together, these quantities distinguish enhanced emission from its interference with the transmitted field: a larger decay rate alone does not imply a deeper transmission dip. The plotted shifts retain the fitted resonant contribution, rather than an absolute dielectric Lamb shift.

\section{Conclusion}
\label{sec: Conclusion}

In this paper, we derived a modal quantum input-output formalism directly from macroscopic QED. In conventional approaches, the eigenmodes of a cavity or waveguide are first identified and then quantized. We reverse the order by first quantizing the complete electromagnetic field under the first-order retarded Green's function framework, and then utilising it to project onto flux-normalized input and output modal channels. The first-order dual Green's function retains both the electric and magnetic fields, allowing the propagation and boundary projections to be expressed compactly. The resulting theory applies to linear, dispersive, and absorptive photonic structures while retaining the full frequency dependence required for both Markovian and non-Markovian dynamics.

When the response is dominated by a single cavity mode or a uniform waveguide, the theory recovers the conventional cavity- and waveguide-QED input-output expressions. More generally, it does not require the interior of the system to have a simple modal description. It can therefore be applied to structures with photonic band edges~\cite{hood2016atom, goban2014}, coupled resonances, and material systems in which loss~\cite{Weiss2018RSE} and dispersion play a central role, such as organic molecular ensembles~\cite{lange2024superradiant}, as well as complex inverse- and machine-learning-designed geometries~\cite{molesky2018inverse, Zhou2026InvDes}. For emitters in the low-saturation regime, or exactly within the single-excitation subspace, the transmission and reflection are determined by the background scattering matrix and Green's functions connecting the input modes to the emitters, the emitters to one another, and the emitters to the output modes.

This provides a direct numerical recipe for quantum photonics: a small set of classical Green-function calculations, obtained for example through FDTD or finite-element simulations, determines the quantum optical response without constructing and quantizing a separate modal model of the device. Future numerical work could test strongly absorbing and dispersive systems, including excitonic materials and magnetodielectrics with negative-index bands, using causal material models and consistent electric and magnetic noise channels~\cite{ciattoni2024quantum, ciattoni2025quantum-optical}. Explicit emitters must be distinguished from excitations already included in the background material response. Extensions to multiphoton scattering and nonlinear emitter dynamics would probe quantum effects beyond the stationary linear response considered here. Extending the treatment to nonlinear material response is another direction motivated by recent quantum descriptions of lossy epsilon-near-zero media~\cite{dalnegro2025nonlinear,poddar2026effective}.

\section{Acknowledgments}

This work is supported by an Energy Frontier Research Center funded by the U.S. Department of Energy (DOE), Office of Science, Basic Energy Sciences (BES), under award DE-SC0025620.

\section{Data Availability}

The data that support the findings of this article are not publicly available. The data are available from the authors upon reasonable request.

\appendix
\newpage
\section{First-Order Waveguide Formalism}
\label{app:dual_modes}

This appendix develops the boundary-mode construction underlying Eqs.~\eqref{eq:flux_normalization}--\eqref{eq:wg_closure} of the main text, making its sign convention and propagating-channel restriction explicit. Following the standard waveguide construction~\cite{Synder&love,svendsen2013reciprocity}, we derive the first-order dual eigenproblem, flux inner product, and modal completeness relation.

As in the main text, the lossless, translationally invariant assumption applies only near each input or output reference plane. The scattering region between the planes may contain arbitrary dispersive, lossy, or structured media.

At a fixed frequency $\omega$, we take the relative permittivity and permeability tensors $\boldsymbol{\varepsilon}(\mathbf{r}_\perp,\omega)$ and $\boldsymbol{\mu}(\mathbf{r}_\perp,\omega)$ to depend only on the transverse coordinates $\mathbf{r}_\perp=(x,y)$. With the $e^{-i\omega t}$ convention, the source-free Maxwell curl equations read
\begin{align}
\nabla \times \mathbf{E}(\mathbf{r}, \omega) &= i\omega\mu_0\,\boldsymbol{\mu}(\mathbf{r}_\perp, \omega)\,\mathbf{H}(\mathbf{r}, \omega),   \nonumber \\
\nabla \times \mathbf{H}(\mathbf{r}, \omega) &= -i\omega\varepsilon_0\,\boldsymbol{\varepsilon}(\mathbf{r}_\perp, \omega)\,\mathbf{E}(\mathbf{r}, \omega).
\label{eq:app_modes_wg_curl_H}
\end{align}
Translational invariance along $z$ gives the modal ansatz of Eq.~\eqref{eq:modal_ansatz}, repeated here for convenience:
\begin{equation}
\mathbf{E}(\mathbf{r}) = \mathbf{E}(\mathbf{r}_\perp)\,e^{i\beta z}, \qquad \mathbf{H}(\mathbf{r}) = \mathbf{H}(\mathbf{r}_\perp)\,e^{i\beta z},
\label{eq:app_modes_modal_ansatz}
\end{equation}
where $\beta$ is the propagation constant to be determined. Under this ansatz, $\partial_z \to i\beta$, the curl operator splits as~\cite{Synder&love}
\begin{equation}
\nabla\times \;\to\; \nabla_t\times + i\beta\,(\hat{\mathbf{z}}\times),
\label{eq:app_modes_curl_split}
\end{equation}
where $\nabla_t\times \equiv \nabla\times$ with $\partial_z \to 0$ is the transverse part of the curl, and $\hat{\mathbf{z}}\times$ is the cross product with $\hat{\mathbf{z}}$, satisfying $(\hat{\mathbf{z}}\times)^2 = -I_t$ with $I_t = \mathbf{1} - \hat{\mathbf{z}}\hat{\mathbf{z}}$ the transverse projector, so it acts as a $\pi/2$ rotation in the cross-section that annihilates longitudinal components.

Substituting Eq.~\eqref{eq:app_modes_curl_split} into Eq.~\eqref{eq:app_modes_wg_curl_H} gives
\begin{align}
\nabla_t \times \mathbf{E} + i\beta\,(\hat{\mathbf{z}} \times \mathbf{E}) &= ik_0 Z_0\,\boldsymbol{\mu}\,\mathbf{H}, \nonumber\\
\nabla_t \times \mathbf{H} + i\beta\,(\hat{\mathbf{z}} \times \mathbf{H}) &= -i\frac{k_0}{Z_0}\,\boldsymbol{\varepsilon}\,\mathbf{E}, \label{eq:app_modes_wg_curl_H_decomp}
\end{align}
Here $k_0=\omega/c$ and $Z_0=\sqrt{\mu_0/\varepsilon_0}$, and the common longitudinal phase $e^{i\beta z}$ has been factored out.

Using the dual-field notation of the main text and Ref.~\cite{partI}, we write $\mathcal{E}(\mathbf{r}_\perp)=[\mathbf{E}(\mathbf{r}_\perp),\;Z_0\mathbf{H}(\mathbf{r}_\perp)]^T$, with $\bar\varepsilon=\operatorname{diag}(\boldsymbol\varepsilon,\boldsymbol\mu)$, and define the dual operators
\begin{align}
\bar{\nabla}_t\times &= \begin{bmatrix} 0 & \nabla_t\times \\ -\nabla_t\times & 0 \end{bmatrix}, \nonumber\\
\bar{\mathbf{z}}\times &= \begin{bmatrix} 0 & \hat{\mathbf{z}}\times \\ -\hat{\mathbf{z}}\times & 0 \end{bmatrix},
\label{eq:app_modes_wg_block_operators}
\end{align}
With these definitions, Eq.~\eqref{eq:app_modes_wg_curl_H_decomp} becomes the first-order eigenproblem on the waveguide cross-section,
\begin{equation}
(i\bar{\nabla}_t\times\ - k_0\bar{\varepsilon})\,\mathcal{E}_n = \beta_n\,(\bar{\mathbf{z}}\times)\,\mathcal{E}_n.
\label{eq:app_modes_wg_eigenproblem_explicit}
\end{equation}
For lossless material, we choose a self-adjoint transverse domain so that the boundary term from integration by parts vanishes: guided modes decay at transverse infinity, while radiation modes can be treated using enclosing self-adjoint boundaries followed by the continuum limit. Defining $\mathcal H_t=i\bar\nabla_t\times$ and $\mathcal M_t=\mathcal H_t-k_0\bar\varepsilon$, the eigenproblem becomes
\begin{equation}
\mathcal{M}_t\,\mathcal{E}_n(\mathbf{r}_\perp) = \beta_n\,(\bar{\mathbf{z}}\times)\,\mathcal{E}_n(\mathbf{r}_\perp).
\label{eq:app_modes_wg_eigenproblem}
\end{equation}
At fixed $\omega$, this generalized eigenproblem determines the propagation constants $\beta_n$ and mode profiles $\mathcal{E}_n(\mathbf{r}_\perp)$, with the longitudinal flux operator $\bar{\mathbf{z}}\times$ as its weight.

Write $B=\bar{\mathbf z}\times$ and $\Pi_t=\operatorname{diag}(I_t,I_t)$. The block signs give $B^\dagger=B$ and $B^2=\Pi_t$. Thus $B$ mixes the tangential electric and magnetic fields and annihilates their longitudinal components. The full mode profile $\mathcal E_n$ may have nonzero longitudinal components; its tangential trace is $\mathcal E_{t,n}=\Pi_t\mathcal E_n$.

The weight in Eq.~\eqref{eq:app_modes_wg_eigenproblem} defines the cross-sectional flux inner product. This is an indefinite Hermitian form: its self-overlap can be positive, negative, or zero.
\begin{equation}
\langle\mathcal{E}_1|\bar{\mathbf{z}}\times\mathcal{E}_2\rangle_S \equiv \int_S \mathcal{E}_1^\dagger(\mathbf{r}_\perp)\,(\bar{\mathbf{z}}\times)\,\mathcal{E}_2(\mathbf{r}_\perp)\,dS.
\label{eq:app_modes_flux_form}
\end{equation}
Expanding in components and using the scalar triple-product identity gives
\begin{multline}
\langle\mathcal{E}_1|\bar{\mathbf{z}}\times\mathcal{E}_2\rangle_S
= -Z_0\!\int_S\! \hat{\mathbf{z}}\cdot\bigl(\mathbf{E}_1^*\!\times\!\mathbf{H}_2
+ \mathbf{E}_2\!\times\!\mathbf{H}_1^*\bigr)\,dS.
\label{eq:app_modes_flux_form_components}
\end{multline}
For a single mode, $\mathcal{E}_1=\mathcal{E}_2=\mathcal{E}$, this reduces to
\begin{equation}
\langle\mathcal{E}|\bar{\mathbf{z}}\times\mathcal{E}\rangle_S
= -4Z_0\!\int_S \hat{\mathbf{z}}\cdot\mathbf{S}_{\rm avg}\,dS,
\label{eq:app_modes_flux_poynting}
\end{equation}
where $\mathbf S_{\rm avg}=\tfrac12\operatorname{Re}(\mathbf E\times\mathbf H^*)$ is the time-averaged Poynting vector. The minus sign follows from the block convention in Eq.~\eqref{eq:app_modes_wg_block_operators}; the form therefore measures minus the longitudinal power flux~\cite{collin1990field,felsen1994radiation}.

For lossless material and the transverse boundary conditions stated above, $\mathcal M_t$ is self-adjoint. Evaluating $\langle\mathcal{E}_m|\mathcal{M}_t\mathcal{E}_n\rangle_S$ with the eigenvalue equation for mode $n$ gives $\beta_n\langle\mathcal{E}_m|\bar{\mathbf{z}}\times\mathcal{E}_n\rangle_S$; moving $\mathcal{M}_t$ onto mode $m$ instead gives $\beta_m^*\langle\mathcal{E}_m|\bar{\mathbf{z}}\times\mathcal{E}_n\rangle_S$.  Subtracting yields the fundamental orthogonality relation~\cite{johnson2002adiabatic, svendsen2013reciprocity}
\begin{equation}
(\beta_n - \beta_m^*)\,\langle\mathcal{E}_m|\bar{\mathbf{z}}\times\mathcal{E}_n\rangle_S = 0.
\label{eq:app_modes_flux_orthogonality}
\end{equation}
For a mode with nonzero self-flux, setting $m=n$ gives $\beta_n=\beta_n^*$. Modes with distinct real propagation constants are flux-orthogonal. We restrict the following normalization to propagating channels with nonzero flux, away from cutoffs, and choose a flux-orthogonal basis within each degenerate subspace:
\begin{equation}
\langle\mathcal{E}_m|\bar{\mathbf{z}}\times\mathcal{E}_n\rangle_S = s_m\,\delta_{mn}, \qquad s_m = \pm 1,
\label{eq:app_modes_flux_normalization}
\end{equation}
With the present convention, $s_m=-\operatorname{sgn}\!\int_S\hat{\mathbf z}\cdot\mathbf S_{{\rm avg},m}\,dS$: forward energy flow has $s_m=-1$ and backward energy flow has $s_m=+1$. For positive-energy modes whose group and energy velocities coincide, $s_m=-\operatorname{sgn}(v_{g,m})$.

For an evanescent mode with nonreal $\beta_n$, Eq.~\eqref{eq:app_modes_flux_orthogonality} instead gives zero self-flux, so the $\pm1$ normalization does not apply. Describing arbitrary near fields requires a generalized modal pairing that includes evanescent modes~\cite{svendsen2013reciprocity}. The input--output theory uses only propagating channels; from this point onward, the mode index $n$ runs over these channels. Radiation channels are included through the corresponding continuum integrals and Dirac normalization.

Any boundary field in the propagating channel basis has the same expansion used in Eq.~\eqref{eq:field_expansion}:
\begin{equation}
\begin{aligned}
\mathcal E(\mathbf r_\perp)
&=\sum_n a_n\mathcal E_n(\mathbf r_\perp),\\
a_n&=s_n\langle\mathcal E_n|\bar{\mathbf z}\times\mathcal E\rangle_S.
\end{aligned}
\label{eq:app_modes_field_expansion}
\end{equation}
The corresponding completeness relation is
\begin{equation}
\sum_n s_n\mathcal E_n(\mathbf r_\perp)
\mathcal E_n^\dagger(\mathbf r'_\perp)
(\bar{\mathbf z}\times)
=\Pi_t\delta^{(2)}(\mathbf r_\perp-\mathbf r'_\perp).
\label{eq:app_modes_wg_closure}
\end{equation}
Equivalently, in bra-ket notation,
\begin{equation}
\sum_n s_n|\mathcal E_n\rangle\langle\mathcal E_n|(\bar{\mathbf z}\times)=\Pi_t,
\label{eq:app_modes_wg_closure_braket}
\end{equation}
where the bras act through integration over the cross-section $S$.
As in the main text, this equality is understood on boundary fields constructed from the propagating input and output channels. It is not a completeness statement for arbitrary near fields, which may also contain evanescent components. These relations provide the propagating-channel basis for Eq.~\eqref{eq:quantum_io}, subject to the sign convention and channel-space restriction specified above.

\section{Canonical output commutator}
\label{app:output_commutator_full}

The goal of this appendix is to prove that the output-mode operators obey the same canonical bosonic commutation relation as the input modes,
\begin{equation}
\left[
\hat b_m(\omega),
\hat b_n^\dagger(\omega')
\right]
=
\delta_{mn}\delta(\omega-\omega').
\label{eq:app_output_commutator_goal}
\end{equation}

We begin with the modal input--output relation in Eq.~\eqref{eq:quantum_io},
\begin{equation}
\hat b_m(\omega)=\sum_p S_{mp}(\omega)\hat a_p(\omega)+\hat\xi_m(\omega),
\label{eq:app_output_io_start}
\end{equation}
and use the bosonic input commutator
$[\hat a_p(\omega),\hat a_q^\dagger(\omega')]
=\delta_{pq}\delta(\omega-\omega')$.
The input operators commute with the independent material-noise operators, so
$[\hat a_p,\hat\xi_n^\dagger]=[\hat\xi_m,\hat a_q^\dagger]=0$.
Expanding the output commutator and dropping these mixed terms gives
\begin{equation}
\begin{aligned}
&[\hat b_m(\omega),\hat b_n^\dagger(\omega')]\\
&\quad=\sum_{p,q}S_{mp}(\omega)S_{nq}^{*}(\omega')
[\hat a_p(\omega),\hat a_q^\dagger(\omega')]\\
&\qquad+[\hat\xi_m(\omega),\hat\xi_n^\dagger(\omega')]\\
&\quad=\sum_p S_{mp}(\omega)S_{np}^{*}(\omega)
\delta(\omega-\omega')\\
&\quad+[\hat\xi_m(\omega),\hat\xi_n^\dagger(\omega')].
\end{aligned}
\label{eq:app_output_scattering_plus_noise}
\end{equation}
Here the frequency delta function allows both scattering amplitudes to be evaluated at $\omega$, and the mode sum is $(SS^\dagger)_{mn}$. The remaining noise commutator is to be evaluated from Eq.~\eqref{eq:noise_operator}.

To establish Eq.~\eqref{eq:app_output_commutator_goal}, the material-noise commutator must supply the contribution missing from $SS^\dagger$. We therefore need to show that
\begin{equation}
[\hat\xi_m(\omega),\hat\xi_n^\dagger(\omega')]
=\bigl[I-SS^\dagger\bigr]_{mn}\delta(\omega-\omega').
\label{eq:app_output_noise_identity_target}
\end{equation}

\subsection{Material-noise contribution}

We choose the integration boundary $\partial V$ in lossless, uniform leads surrounding the scattering region. The interior material may be spatially varying, dispersive, and absorbing.

Writing Eq.~\eqref{eq:noise_operator} explicitly,
\begin{equation}
\begin{aligned}
&\hat\xi_m(\omega)\\
&\quad=\alpha k_0
\int_{S_{\rm out}}dS_{\mathbf r}\int_Vd^3\mathbf x\;
\mathcal E_m^{{\rm out}\dagger}(\mathbf r,\omega)\\
&\qquad\times(\bar n_{\mathbf r}\times)
g(\mathbf r,\mathbf x;\omega)
\hat{\mathcal P}_N(\mathbf x,\omega),
\end{aligned}
\label{eq:app_noise_operator_braket}
\end{equation}
where $\alpha=\sqrt{2\pi\varepsilon_0/(\hbar k_0)}$ and all boundary normals point outward from $V$. The polarization-noise commutator, Eq.~\eqref{eq:PN_commutator}, is
\begin{equation*}
\begin{aligned}
&[\hat{\mathcal P}_{N,i}(\mathbf x,\omega),
\hat{\mathcal P}_{N,j}^\dagger(\mathbf y,\omega')]\\
&\quad=\frac{\hbar}{\pi\varepsilon_0}
(\bar\varepsilon_I)_{ij}(\mathbf x,\omega)
\delta^{(3)}(\mathbf x-\mathbf y)\delta(\omega-\omega').
\end{aligned}
\end{equation*}
Substitution collapses the two source-volume integrals to one:
\begin{equation}
\begin{aligned}
&[\hat\xi_m(\omega),\hat\xi_n^\dagger(\omega')]\\
&=2k_0\delta(\omega-\omega')
\int_{S_{\rm out}}dS_{\mathbf r}
\int_{S_{\rm out}}dS_{\mathbf r'}
\int_Vd^3\mathbf x\\
&\quad\times
\mathcal E_m^{{\rm out}\dagger}(\mathbf r)
(\bar n_{\mathbf r}\times)g(\mathbf r,\mathbf x)
\bar\varepsilon_I(\mathbf x)\\
&\quad\times
g^\dagger(\mathbf r',\mathbf x)
(\bar n_{\mathbf r'}\times)\mathcal E_n^{\rm out}(\mathbf r').
\end{aligned}
\label{eq:app_noise_green_overlap}
\end{equation}
All classical quantities are now evaluated at $\omega$, with that argument suppressed. Here $g^\dagger(\mathbf r,\mathbf x)$ denotes the matrix Hermitian conjugate of $g(\mathbf r,\mathbf x)$. We used $(\bar n\times)^\dagger=(\bar n\times)$ and $\alpha^2k_0^2\hbar/(\pi\varepsilon_0)=2k_0$. The volume integral includes absorption throughout the interior.


We evaluate the absorption overlap by expressing it as the boundary flux of auxiliary fields. Their projections onto incoming and outgoing port modes then give the scattering contribution and the identity contribution, respectively.

The detector planes $S_{\rm out}$ are placed just outside the integration boundary so that the auxiliary fields have no sources within $V$. Define
\begin{equation}
\psi_n(\mathbf x)
=\int_{S_{\rm out}}dS_{\mathbf r}\;
g^\dagger(\mathbf r,\mathbf x)
(\bar n_{\mathbf r}\times)\mathcal E_n^{\rm out}(\mathbf r).
\label{eq:app_final_auxiliary_field}
\end{equation}
Because $\psi_n$ is constructed with $g^\dagger$, it obeys the adjoint Maxwell equation. It is not the field produced by launching mode $n$ into the original absorbing medium. Its Hermitian conjugate is
\begin{equation}
\psi_m^\dagger(\mathbf x)
=\int_{S_{\rm out}}dS_{\mathbf r}\;
\mathcal E_m^{{\rm out}\dagger}(\mathbf r)
(\bar n_{\mathbf r}\times)g(\mathbf r,\mathbf x).
\label{eq:app_final_auxiliary_bra}
\end{equation}
These are precisely the factors on either side of $\bar\varepsilon_I$ in Eq.~\eqref{eq:app_noise_green_overlap}, giving
\begin{equation}
\begin{aligned}
&[\hat\xi_m(\omega),\hat\xi_n^\dagger(\omega')]\\
&\quad=2k_0\delta(\omega-\omega')
\int_Vd^3\mathbf x\;
\psi_m^\dagger(\mathbf x)\bar\varepsilon_I(\mathbf x)\psi_n(\mathbf x).
\end{aligned}
\label{eq:app_final_noise_volume}
\end{equation}

Since the detector sources lie outside $V$, the interior equation is
\begin{equation}
(\mathcal H-k_0\bar\varepsilon^\dagger)\psi_n=0.
\label{eq:app_final_adjoint_maxwell}
\end{equation}
Thus, for the original Maxwell operator $\mathcal M=\mathcal H-k_0\bar\varepsilon$,
\begin{equation}
\begin{aligned}
\mathcal M\psi_n&=-2ik_0\bar\varepsilon_I\psi_n,\\
(\mathcal M\psi_m)^\dagger
&=2ik_0\psi_m^\dagger\bar\varepsilon_I.
\end{aligned}
\label{eq:app_final_maxwell_action}
\end{equation}
In particular, $\mathcal M\psi_n$ need not vanish in the absorbing interior.

The Maxwell integration-by-parts identity from Ref.~\cite{partI} reads
\begin{equation}
\begin{aligned}
&\int_Vd^3\mathbf x\;
\left[\psi_m^\dagger(\mathcal M\psi_n)
-(\mathcal M\psi_m)^\dagger\psi_n\right]\\
&\quad=-2ik_0\int_Vd^3\mathbf x\;
\psi_m^\dagger\bar\varepsilon_I\psi_n
+i\int_{\partial V}dS\;
\psi_m^\dagger(\bar n\times)\psi_n.
\end{aligned}
\label{eq:app_final_maxwell_identity}
\end{equation}
By Eq.~\eqref{eq:app_final_maxwell_action}, the left-hand side is $-4ik_0\int_V\psi_m^\dagger\bar\varepsilon_I\psi_n\,d^3\mathbf x$. Substituting and rearranging therefore gives
\begin{equation}
\begin{aligned}
&2k_0\int_Vd^3\mathbf x\;
\psi_m^\dagger\bar\varepsilon_I\psi_n
\\
&\qquad=-\int_{\partial V}dS\;
\psi_m^\dagger(\bar n\times)\psi_n.
\end{aligned}
\label{eq:app_final_volume_to_flux}
\end{equation}
The noise commutator becomes
\begin{equation}
\begin{aligned}
&[\hat\xi_m(\omega),\hat\xi_n^\dagger(\omega')]\\
&\quad=-\delta(\omega-\omega')
\int_{\partial V}dS_{\mathbf x}\;
\psi_m^\dagger(\mathbf x)
(\bar n_{\mathbf x}\times)\psi_n(\mathbf x).
\end{aligned}
\label{eq:app_final_noise_surface}
\end{equation}

\subsection{Evaluation in the port basis}

Input reference planes lie on $\partial V$, and outgoing modes are transported to the detector planes through the lossless leads with their phases retained. The boundary modes obey the flux normalization of Appendix~\ref{app:dual_modes}. With outward normals,
\begin{equation}
\begin{aligned}
\int_{\partial V}dS\;
\mathcal E_p^{{\rm in}\dagger}(\bar n\times)\mathcal E_q^{\rm in}
&=\delta_{pq},\\
\int_{\partial V}dS\;
\mathcal E_p^{{\rm out}\dagger}(\bar n\times)\mathcal E_q^{\rm out}
&=-\delta_{pq},\\
\int_{\partial V}dS\;
\mathcal E_p^{{\rm in}\dagger}(\bar n\times)\mathcal E_q^{\rm out}
&=0.
\end{aligned}
\label{eq:app_final_channel_norms}
\end{equation}
Here the incoming and outgoing labels denote propagation directions on each port; both directions must be retained on the complete boundary.

Projecting Eq.~\eqref{eq:app_final_auxiliary_field} onto an incoming mode and conjugating the scalar overlap gives
\begin{equation}
\begin{aligned}
&\int_{\partial V}dS_{\mathbf x}\;
\mathcal E_p^{{\rm in}\dagger}(\mathbf x)
(\bar n_{\mathbf x}\times)\psi_n(\mathbf x)\\
&=\Bigg[
\int_{S_{\rm out}}dS_{\mathbf r}\int_{\partial V}dS_{\mathbf x}\;
\mathcal E_n^{{\rm out}\dagger}(\mathbf r)
(\bar n_{\mathbf r}\times)\\
&\qquad\qquad\times g(\mathbf r,\mathbf x)
(\bar n_{\mathbf x}\times)\mathcal E_p^{\rm in}(\mathbf x)
\Bigg]^*\\
&=(iS_{np})^*=-iS_{np}^*.
\end{aligned}
\label{eq:app_final_incoming_projection}
\end{equation}
The last step uses the definition of $S_{np}$ in Eq.~\eqref{eq:S-matrix}, with the input and detector reference planes specified above.

As for the incoming projection, we conjugate the outgoing overlap so that it involves the retarded Green function. An outgoing equivalent surface source then propagates directly to the detector plane, where mode orthogonality fixes its amplitude. Specifically, the exterior boundary-field representation of Ref.~\cite{partI} states that equivalent electric and magnetic surface currents constructed from an outgoing mode reproduce that mode outside the boundary and give zero field inside. In the present convention,
\begin{equation}
\begin{aligned}
&i\int_{\partial V}dS_{\mathbf x}\;
g(\mathbf r,\mathbf x)(\bar n_{\mathbf x}\times)
\mathcal E_p^{\rm out}(\mathbf x)\\
&\qquad=\mathcal E_p^{\rm out}(\mathbf r),
\quad \mathbf r\in S_{\rm out}.
\end{aligned}
\label{eq:app_final_outward_propagation}
\end{equation}
The sign follows by extending the outgoing field by zero into $V$: the curl across the boundary produces the source $i(\bar n\times)\mathcal E_p^{\rm out}\delta_{\partial V}$, which is propagated by $g$. This uses the full retarded Green function; the interior material does not affect this source's response because its interior field is zero.

Using Eq.~\eqref{eq:app_final_outward_propagation} in the conjugated outgoing overlap gives
\begin{equation}
\begin{aligned}
&\int_{\partial V}dS_{\mathbf x}\;
\mathcal E_p^{{\rm out}\dagger}(\mathbf x)
(\bar n_{\mathbf x}\times)\psi_n(\mathbf x)\\
&=\Bigg[
\int_{S_{\rm out}}dS_{\mathbf r}\;
\mathcal E_n^{{\rm out}\dagger}(\mathbf r)
(\bar n_{\mathbf r}\times)\\
&\qquad\times\int_{\partial V}dS_{\mathbf x}\;
g(\mathbf r,\mathbf x)(\bar n_{\mathbf x}\times)
\mathcal E_p^{\rm out}(\mathbf x)
\Bigg]^*\\
&=\left[
-i\int_{S_{\rm out}}dS_{\mathbf r}\;
\mathcal E_n^{{\rm out}\dagger}(\mathbf r)
(\bar n_{\mathbf r}\times)\mathcal E_p^{\rm out}(\mathbf r)
\right]^*\\
&=(i\delta_{np})^*=-i\delta_{pn}.
\end{aligned}
\label{eq:app_final_outgoing_projection}
\end{equation}
Because the outgoing flux norm is negative, the outgoing expansion coefficient is $i\delta_{pn}$. Equations~\eqref{eq:app_final_incoming_projection} and~\eqref{eq:app_final_outgoing_projection} therefore determine the propagating part of the boundary field:
\begin{equation}
\psi_n\big|_{\partial V}
=i\mathcal E_n^{\rm out}
-i\sum_p S_{np}^*\mathcal E_p^{\rm in}.
\label{eq:app_final_boundary_expansion}
\end{equation}
The channel basis includes all reflected, transmitted, and radiated propagating channels, with continuum sums understood as integrals. The flux is evaluated on asymptotic lossless lead sections, where decaying evanescent fields do not contribute, and no exterior scattering or feedback is present. Only boundary modes have been expanded; the interior remains described by its full Green function.

Applying Eq.~\eqref{eq:app_final_channel_norms} now yields
\begin{equation}
\begin{aligned}
\int_{\partial V}dS\;
\psi_m^\dagger(\bar n\times)\psi_n
&=(-i)(i)(-\delta_{mn})\\
&\quad+\sum_p(iS_{mp})(-iS_{np}^*)\\
&=-\delta_{mn}+(SS^\dagger)_{mn}.
\end{aligned}
\label{eq:app_final_boundary_flux}
\end{equation}
Substitution into Eq.~\eqref{eq:app_final_noise_surface} proves
\begin{equation}
[\hat\xi_m(\omega),\hat\xi_n^\dagger(\omega')]
=\bigl[I-SS^\dagger\bigr]_{mn}\delta(\omega-\omega').
\label{eq:app_final_noise_identity}
\end{equation}
Together with Eq.~\eqref{eq:app_output_scattering_plus_noise}, this gives
\begin{equation}
\begin{aligned}
&[\hat b_m(\omega),\hat b_n^\dagger(\omega')]\\
&\quad=\bigl[SS^\dagger+I-SS^\dagger\bigr]_{mn}
\delta(\omega-\omega')\\
&\quad=\delta_{mn}\delta(\omega-\omega'),
\end{aligned}
\label{eq:app_final_output_identity}
\end{equation}
as required.

For passive material, the volume overlap in Eq.~\eqref{eq:app_final_noise_volume} is positive semidefinite, so the material noise supplies exactly the deficit $I-SS^\dagger$, including at zero temperature. With no material absorption, that overlap vanishes and the complete channel relation reduces to $SS^\dagger=I$. A restricted transmission matrix can remain nonunitary because reflection or radiation into omitted channels also removes transmitted power. No reciprocity assumption is required in this proof.

\section{Cavity eigenmode expansion and single-resonance limit}
\label{app:cavity_green_derivation}

We derive the cavity Green-function expansion by constructing an operator that satisfies $\mathcal M G=I$. Consider a closed, lossless system with a nondispersive, Hermitian, positive-definite material tensor $\bar\varepsilon(\mathbf r)$. The boundary conditions are chosen so that the bilinear boundary flux vanishes, making $\mathcal H=i\bar\nabla\times$ self-adjoint. The source-free modes satisfy the generalized eigenvalue equation~\cite{sakoda2004optical,silveirinha2024modal}
\begin{equation}
\mathcal H|\mathcal E_n\rangle
=k_{0,n}\bar\varepsilon|\mathcal E_n\rangle,
\qquad k_{0,n}=\omega_n/c.
\label{eq:cgd_eigenproblem}
\end{equation}
Self-adjointness gives $\langle\mathcal E_m|\mathcal H\mathcal E_n\rangle=\langle\mathcal H\mathcal E_m|\mathcal E_n\rangle$. Substituting Eq.~\eqref{eq:cgd_eigenproblem} on both sides yields
\begin{equation}
(k_{0,n}-k_{0,m}^*)
\langle\mathcal E_m|\bar\varepsilon|\mathcal E_n\rangle=0.
\label{eq:cgd_orthogonality}
\end{equation}
Since $\bar\varepsilon$ is positive definite, setting $m=n$ establishes that $k_{0,n}$ is real. Modes with distinct eigenvalues are therefore orthogonal under the material-weighted inner product. Choosing an orthogonal basis within each degenerate subspace, we normalize the modes as
\begin{equation}
\langle\mathcal E_m|\bar\varepsilon|\mathcal E_n\rangle
=\int_V\mathcal E_m^\dagger\bar\varepsilon\mathcal E_n\,dV
=\delta_{mn}.
\label{eq:cgd_normalization}
\end{equation}
Here the inner product is weighted by $\bar\varepsilon$ and integrated over the volume, so the norm measures stored electromagnetic energy. The choice of $\bar{\varepsilon}$ weighing arises to solve for the standing wave eigenfrequencies, whereas the $\bar{\mathbf z} \times $ weighing in Appendix~\ref{app:dual_modes} solves for propagating modes at fixed frequency.

Absorbing the material weight into the field,
\begin{equation}
|\mathcal F_n\rangle=\bar\varepsilon^{1/2}|\mathcal E_n\rangle,
\qquad
\mathcal K=\bar\varepsilon^{-1/2}\mathcal H\bar\varepsilon^{-1/2},
\label{eq:cgd_transform_definitions}
\end{equation}
then multiplying Eq.~\eqref{eq:cgd_eigenproblem} by $\bar\varepsilon^{-1/2}$ gives
\begin{equation}
\mathcal K|\mathcal F_n\rangle=k_{0,n}|\mathcal F_n\rangle,
\qquad
\langle\mathcal F_m|\mathcal F_n\rangle=\delta_{mn}.
\label{eq:cgd_Hermitian_transform}
\end{equation}
For the closed-system spectral problem considered here, the modes of this self-adjoint operator form a complete basis and their ordinary completeness relation is
\begin{equation}
\sum_n|\mathcal F_n\rangle\langle\mathcal F_n|=I.
\label{eq:cgd_transformed_closure}
\end{equation}
Substituting Eq.~\eqref{eq:cgd_transform_definitions} and multiplying on both sides by $\bar\varepsilon^{-1/2}$ gives
\begin{equation}
\sum_n|\mathcal E_n\rangle\langle\mathcal E_n|
=\bar\varepsilon^{-1}.
\label{eq:cgd_unweighted_sum}
\end{equation}
Consequently,
\begin{equation}
\begin{aligned}
\sum_n\bar\varepsilon|\mathcal E_n\rangle\langle\mathcal E_n|&=I\ \text{or}\ 
\sum_n|\mathcal E_n\rangle\langle\mathcal E_n|\bar\varepsilon&=I.
\end{aligned}
\label{eq:cgd_closure}
\end{equation}
These identities refer to the complete field space being inverted, including negative-frequency modes and any required static sector. If the problem is restricted to a constrained subspace, its completeness relation and inverse must be restricted consistently.

We can now construct the Green operator directly. Acting on an eigenmode,
\begin{equation}
\mathcal M|\mathcal E_n\rangle
=(k_{0,n}-k_0)\bar\varepsilon|\mathcal E_n\rangle.
\label{eq:cgd_M_on_mode}
\end{equation}
For $k_0$ outside the spectrum, consider
\begin{equation}
G=\sum_n\frac{|\mathcal E_n\rangle\langle\mathcal E_n|}
{k_{0,n}-k_0}.
\label{eq:cgd_inverse_candidate}
\end{equation}
Applying $\mathcal M$ and using Eqs.~\eqref{eq:cgd_M_on_mode} and~\eqref{eq:cgd_closure},
\begin{equation}
\begin{aligned}
\mathcal M G
&=\sum_n\frac{\mathcal M|\mathcal E_n\rangle\langle\mathcal E_n|}
{k_{0,n}-k_0}\\
&=\sum_n\bar\varepsilon|\mathcal E_n\rangle\langle\mathcal E_n|=I.
\end{aligned}
\label{eq:cgd_inverse_check}
\end{equation}
Similarly, $\langle\mathcal E_n|\mathcal M=(k_{0,n}-k_0)\langle\mathcal E_n|\bar\varepsilon$ gives $G\mathcal M=I$ on the operator domain. Equation~\eqref{eq:cgd_inverse_candidate} is therefore the inverse Maxwell operator. With the $e^{-i\omega t}$ convention, the retarded Green function follows by approaching the real frequency from $\omega+i0^+$~\cite{novotny2006principles}:
\begin{equation}
G(\omega)=\sum_n\frac{|\mathcal E_n\rangle\langle\mathcal E_n|}
{k_{0,n}-k_0-i0^+}.
\label{eq:cgd_resolvent}
\end{equation}
In position space,
\begin{equation}
g(\mathbf r,\mathbf r',\omega)
=\sum_n\frac{\mathcal E_n(\mathbf r)\mathcal E_n^\dagger(\mathbf r')}
{k_{0,n}-k_0-i0^+}.
\label{eq:cgd_kernel}
\end{equation}
This is Eq.~\eqref{eq:cav_green_spectral}. No additional $\bar\varepsilon$ appears in the numerator: the material weight is already included in the normalization and completeness relations. More generally, electromagnetic energy in absorbing and dispersive dielectrics includes material and reservoir contributions~\cite{rosa2010electromagnetic,rosa2011electromagnetic}; a dispersive eigenmode treatment therefore requires a modified energy normalization or an enlarged field--material eigenproblem. 

For a weakly perturbed cavity with a well-isolated resonance, we retain its closed-system mode profile to leading order and replace the real resonance frequency by $\tilde\omega_c=\omega_c-i\kappa/2$~\cite{kristensen2014quasinormal}. Here $\kappa$ is the total energy-decay rate from radiation and absorption, and $\omega_c$ may include a small real frequency shift. The resonant contribution is then
\begin{equation}
g_{\rm cav}(\mathbf r,\mathbf r',\omega)
\simeq\frac{\mathcal E_c(\mathbf r)\mathcal E_c^\dagger(\mathbf r')}
{k_{0,c}-k_0-i\kappa/(2c)}.
\label{eq:cgd_single_k}
\end{equation}
Writing $\Delta(\omega)=\omega-\omega_c$, the denominator becomes
\begin{equation}
k_{0,c}-k_0-\frac{i\kappa}{2c}
=-\frac{\Delta(\omega)+i\kappa/2}{c},
\label{eq:cgd_conversion}
\end{equation}
so that
\begin{equation}
g_{\rm cav}(\mathbf r,\mathbf r',\omega)
\simeq-\frac{c\,\mathcal E_c(\mathbf r)\mathcal E_c^\dagger(\mathbf r')}
{\Delta(\omega)+i\kappa/2}.
\label{eq:cgd_single_omega}
\end{equation}
The factor $-c$ follows from converting the denominator in momentum space to frequency.

\section{Uniform lossless waveguide Green function}
\label{app:waveguide_green}

We obtain the propagating waveguide Green function from the spectral expansion in Appendix~\ref{app:cavity_green_derivation}, using the eigenmode-expansion and contour-integration approach of Ref.~\cite{sakoda2004optical} in the first-order dual-field formalism. The material tensor is lossless, nondispersive, Hermitian, and positive definite, and is independent of $z$, but may vary across the transverse plane.

Appendix~\ref{app:dual_modes} constructs modes at fixed frequency that are orthogonal under a surface flux pairing, which measures power flow through the cross-section. Here we instead use the frequency eigenmodes of the volume problem, whose orthogonality is defined by the $\bar\varepsilon$-weighted energy inner product of Appendix~\ref{app:cavity_green_derivation}. These are the same physical waveguide solutions, but with a different normalization: the profiles $\mathbf u_{n\beta}$ below are energy-normalized, rather than unit-flux-normalized. The transverse energy normalization follows by separating the longitudinal dependence of the volume modes; it does not replace the fixed-frequency flux orthogonality used in Appendix~\ref{app:dual_modes}.

We first impose periodic boundary conditions over a length $L$, then take $L\to\infty$. The modes can be written as
\begin{equation}
\mathcal E_{n\beta}(\mathbf r)
=L^{-1/2}\mathbf u_{n\beta}(\mathbf r_\perp)e^{i\beta z},
\qquad \beta=2\pi j/L,
\label{eq:wgd_energy_modes}
\end{equation}
where $n$ labels the transverse branch and $\omega_n(\beta)$ is its eigenfrequency. We choose the transverse energy normalization
\begin{equation}
\langle\mathbf u_{n\beta}|\bar\varepsilon|\mathbf u_{n\beta}\rangle_S
=\int_S\mathbf u_{n\beta}^\dagger\bar\varepsilon\mathbf u_{n\beta}\,dS=1,
\label{eq:wgd_energy_norm}
\end{equation}
so that the full mode has the volume normalization of Eq.~\eqref{eq:cgd_normalization}.

Substituting Eq.~\eqref{eq:wgd_energy_modes} into Eq.~\eqref{eq:cgd_kernel} gives
\begin{equation}
g(\mathbf r,\mathbf r',\omega)
=\frac{c}{L}\sum_{n,\beta}
\frac{\mathbf u_{n\beta}(\mathbf r_\perp)\mathbf u_{n\beta}^\dagger(\mathbf r'_\perp)e^{i\beta(z-z')}}
{\omega_n(\beta)-\omega-i0^+}.
\label{eq:wgd_finite_spectral}
\end{equation}
The factor $c$ converts the spectral denominator from wavenumber to frequency. As $L\to\infty$, $L^{-1}\sum_\beta\to\int d\beta/(2\pi)$, yielding the spectral integral
\begin{equation}
g(\mathbf r,\mathbf r',\omega)
=c\sum_n\int\frac{d\beta}{2\pi}
\frac{\mathbf u_{n\beta}(\mathbf r_\perp)\mathbf u_{n\beta}^\dagger(\mathbf r'_\perp)e^{i\beta(z-z')}}
{\omega_n(\beta)-\omega-i0^+}.
\label{eq:wgd_frequency_integral}
\end{equation}
The full spectral representation includes the radiation continuum and all spectral sectors required in Appendix~\ref{app:cavity_green_derivation}; sums over continuous transverse labels are understood as integrals. We now extract the contributions of simple propagating poles, away from cutoffs.

Let $\beta_j$ be a real root of $\omega_n(\beta_j)=\omega$, and let $j$ label the branch and root, including both propagation directions. Near this root,
\begin{equation}
\begin{aligned}
\omega_n(\beta)-\omega-i0^+
&\simeq v_{g,j}(\beta-\beta_j)-i0^+,\\
v_{g,j}&=\left.\frac{d\omega_n}{d\beta}\right|_{\beta_j}\ne0.
\end{aligned}
\label{eq:wgd_pole_linearization}
\end{equation}
The retarded pole lies above the real $\beta$ axis for $v_{g,j}>0$ and below it for $v_{g,j}<0$. Writing $Z=z-z'$, the pole contribution follows from
\begin{equation}
\int\frac{d\beta}{2\pi}
\frac{e^{i\beta Z}}{v_{g,j}(\beta-\beta_j)-i0^+}
=\frac{i}{|v_{g,j}|}e^{i\beta_j Z}\Theta(v_{g,j}Z).
\label{eq:wgd_velocity_residue}
\end{equation}
For $Z>0$ the contour closes in the upper half-plane; for $Z<0$ it closes in the lower half-plane, with the opposite orientation. Each pole therefore contributes on the side toward which its mode carries energy. Evaluating the mode profiles at the pole, $\mathbf u_j=\mathbf u_{n\beta_j}$, gives
\begin{multline}
g_{\rm wg}(\mathbf r,\mathbf r',\omega)
=i\sum_j\frac{c}{|v_{g,j}|}
\mathbf u_j(\mathbf r_\perp)\mathbf u_j^\dagger(\mathbf r'_\perp)\\
\times e^{i\beta_j(z-z')}\Theta\!\bigl(v_{g,j}(z-z')\bigr).
\label{eq:wgd_energy_kernel}
\end{multline}
This expression gives the propagating-pole contribution, not any additional evanescent or longitudinal/contact terms needed for the full near-field Green function.

The energy-normalized profiles $\mathbf u_j$ are related to the unit-flux modes of Appendix~\ref{app:dual_modes} by $\mathcal E_j=\sqrt{c/|v_{g,j}|}\,\mathbf u_j$. For the lossless nondispersive guide considered here, power and energy per unit length obey $P_j=v_{g,j}U_j$; with $\hat{\mathbf n}=\hat{\mathbf z}$, this gives $\langle\mathbf u_j|(\bar n\times)|\mathbf u_j\rangle_S=-v_{g,j}/c$ when $\langle\mathbf u_j|\bar\varepsilon|\mathbf u_j\rangle_S=1$. Thus the explicit factor $c/|v_{g,j}|$ in Eq.~\eqref{eq:wgd_energy_kernel} converts an energy-normalized mode dyad into a flux-normalized one. The two orthogonality conditions refer to different spectral problems: frequency eigenmodes are orthogonal under the volume energy inner product containing $\bar\varepsilon$, whereas the fixed-frequency propagating channels satisfy $\langle\mathcal E_m|(\bar n\times)|\mathcal E_n\rangle_S=s_n\delta_{mn}$, with $s_n=-\operatorname{sgn}(v_{g,n})$. Energy orthogonality does not imply that channels with different propagation constants at the same frequency are orthogonal under a transverse $\bar\varepsilon$-weighted pairing. We retain the energy-normalized profiles and the explicit group-velocity factor here.

\section{Green's functions for Composite Systems}
\label{app:composite_moved}

We construct the Green function of a coupled system by expanding it in cavity and waveguide modes, then finding the expansion coefficients from the Maxwell equation. This gives a two-mode matrix for coupled cavities and a discrete mode coupled to a continuum for a cavity beside a waveguide. Throughout, $g$ inverts $\mathcal M$, so $\mathcal M g=I$, with the normalization of Appendix~\ref{app:cavity_green_derivation}.

The modes below form a retained, energy-orthonormal basis with a common positive, lossless, non-dispersive material weight $\bar\varepsilon$. Thus $\langle\mathcal E_\mu|\bar\varepsilon|\mathcal E_\nu\rangle=\delta_{\mu\nu}$, with a Dirac delta for continuum indices. Isolated-component modes can approximate this basis when their overlap is small; otherwise they must be orthogonalized or their overlap matrix retained. The expansion is a reduced description, not a completeness assertion for the full open structure. Weak leakage into omitted channels is introduced through linewidths after projection.

\subsection{Coupled Cavities}
Consider two cavities whose response is dominated by localized modes $\mathcal E_1$ and $\mathcal E_2$. We choose an energy-orthonormal basis, $\langle\mathcal E_i|\bar\varepsilon|\mathcal E_j\rangle=\delta_{ij}$, where $\bar\varepsilon$ describes the lossless, nondispersive composite structure and the inner product is over the volume, as in Appendix~\ref{app:cavity_green_derivation}. For weakly overlapping cavities, these can be approximated by the normalized isolated-cavity modes. The field is expanded as
\begin{equation}
|\mathcal E\rangle=a_1|\mathcal E_1\rangle+a_2|\mathcal E_2\rangle.
\label{eq:cc_field_expansion}
\end{equation}
Here $a_1$ and $a_2$ are the amplitudes in the two cavities. We use the Maxwell operator $\mathcal M=\mathcal H-(\omega/c)\bar\varepsilon$ of Appendix~\ref{app:cavity_green_derivation}, where $\omega$ is the driving angular frequency and $c$ is the speed of light in vacuum. Multiplying its matrix elements by $c$ converts them from inverse length to angular frequency. We therefore define the projected cavity frequencies $\omega_i$ and coupling rate $J$ as
\begin{equation}
\begin{aligned}
\omega_i&=c\langle\mathcal E_i|\mathcal H|\mathcal E_i\rangle,\\
J&=c\langle\mathcal E_1|\mathcal H|\mathcal E_2\rangle.
\end{aligned}
\label{eq:cc_parameters}
\end{equation}
The frequencies approach those of the isolated cavities as their interaction vanishes. The coupling $J$ is a matrix element of the Maxwell curl operator, not simply the overlap of the two fields. Self-adjointness gives the reverse coupling as $J^*$.

Applying $\mathcal M$ to Eq.~\eqref{eq:cc_field_expansion} and projecting onto each basis mode now gives
\begin{equation}
\begin{aligned}
c\langle\mathcal E_1|\mathcal M|\mathcal E\rangle
&=(\omega_1-\omega)a_1+Ja_2,\\
c\langle\mathcal E_2|\mathcal M|\mathcal E\rangle
&=J^*a_1+(\omega_2-\omega)a_2.
\end{aligned}
\label{eq:cc_projected_equations}
\end{equation}
These are the two coupled amplitude equations. Their coefficient matrix is
\begin{equation}
\mathbf K(\omega)=\begin{pmatrix}
\omega_1-\omega&J\\
J^*&\omega_2-\omega
\end{pmatrix}.
\label{eq:cc_projected_matrix}
\end{equation}

Inverting this matrix determines the amplitudes from the projections of $\mathcal M|\mathcal E\rangle$. Substituting them back into Eq.~\eqref{eq:cc_field_expansion} therefore gives the approximate inverse of $\mathcal M$ in the retained two-mode subspace:
\begin{equation}
G_{\rm cav}\simeq c\sum_{i,j=1}^2
|\mathcal E_i\rangle[\mathbf K^{-1}]_{ij}\langle\mathcal E_j|,
\label{eq:cc_green_operator}
\end{equation}
with the retarded prescription $\omega\to\omega+i0^+$. The factor $c$ appears because $\mathbf K$ represents $c\mathcal M$, whereas the Green operator inverts $\mathcal M$.

To include weak leakage and absorption, let $\kappa_i$ denote the energy-decay rate of cavity $i$, so its field amplitude decays at rate $\kappa_i/2$. As in the single-cavity approximation, replace $\omega_i$ by $\omega_i-i\kappa_i/2$. Assuming independent decay and neglecting damping mediated by shared channels, the matrix becomes
\begin{equation}
\mathbf K(\omega)=\begin{pmatrix}
\omega_1-\omega-i\kappa_1/2&J\\
J^*&\omega_2-\omega-i\kappa_2/2
\end{pmatrix}.
\label{eq:cc_damped_matrix}
\end{equation}
Its determinant sets the coupled resonance denominator,
\begin{equation}
\begin{aligned}
\det\mathbf K={}&(\omega_1-\omega-i\kappa_1/2)\\
&\times(\omega_2-\omega-i\kappa_2/2)-|J|^2,
\end{aligned}
\label{eq:cc_determinant}
\end{equation}
and the inverse is
\begin{equation}
\mathbf K^{-1}=\frac{1}{\det\mathbf K}
\begin{pmatrix}
\omega_2-\omega-i\kappa_2/2&-J\\
-J^*&\omega_1-\omega-i\kappa_1/2
\end{pmatrix},
\label{eq:cc_inverse}
\end{equation}
and therefore
\begin{equation}
\begin{aligned}
G_{\rm cav}\simeq\frac{c}{\det\mathbf K}\Big[&
(\omega_2-\omega-i\kappa_2/2)|\mathcal E_1\rangle\langle\mathcal E_1|\\
&+(\omega_1-\omega-i\kappa_1/2)|\mathcal E_2\rangle\langle\mathcal E_2|\\
&-J|\mathcal E_1\rangle\langle\mathcal E_2|\\
&-J^*|\mathcal E_2\rangle\langle\mathcal E_1|\Big].
\end{aligned}
\label{eq:cc_green_explicit}
\end{equation}
The diagonal terms describe the response within each cavity, while the cross terms describe transfer between them. For $J=0$, the result reduces to the two independent cavity Green functions. For identical lossless cavities, $\omega_1=\omega_2=\omega_c$ and $\kappa_1=\kappa_2=0$, the two poles occur at $\omega_c\pm|J|$, corresponding to the symmetric and antisymmetric combinations after choosing the relative mode phase. Inserting the kernel of Eq.~\eqref{eq:cc_green_explicit} into Eq.~\eqref{eq:S-matrix} then gives the cavity-mediated input--output response; any direct nonresonant path is added separately.

\subsection{A cavity coupled to a waveguide}

For the cavity-waveguide case, the waveguide modes are taken as propagating modes indexed by $k$, including the relevant propagation directions and bands, the field description,
\begin{equation}
|\mathcal E\rangle=a_c|\mathcal E_c\rangle
+\sum_k a_k|\mathcal E_k\rangle.
\label{eq:tmpcomp_continuum_field}
\end{equation}
We demand that this field relation must be describable using the Green's function such that, for the composite system, the Green's function should satisfy,
\begin{equation}
\Big(i\bar{\nabla}\times\ -\frac{\omega}{c}\bar{\varepsilon}_t
(\mathbf{r})\Big) g(\mathbf{r}, \mathbf{r}', \omega) = \frac{\omega}{c}\mathbb{I}_6\delta(\mathbf{r}-\mathbf{r}')
\end{equation}
where the total permittivity can be defined in accordance to Lippmann-Schwinger formulation~\cite{Hughes2004SingleQD, Chen2019GENOME}. 
\begin{equation}
\bar\varepsilon_t(\mathbf r) = \bar{\varepsilon}_b(\mathbf{r}) + \Delta\bar{\varepsilon}_w(\mathbf{r}) + \Delta\bar{\varepsilon}_c(\mathbf r)
\end{equation}
with the waveguide, $\Delta\bar{\varepsilon}_w$, and cavity, $\Delta\bar{\varepsilon}_c$, permittivities being a perturbation in each other's dielectric environment, and $\bar{\varepsilon}_b$ is the background permittivity in the absence of waveguide and cavity.

Under quasi-normal approximation with no cross-coupling between the waveguide modes, one can decompose the GF into the spectral representation as, 
\begin{align}
g(\mathbf r, \mathbf r' , \omega) = \sum_{\{\alpha, \beta\} = \{k, c\}} A_{\alpha, \beta}(\omega)\  \mathcal{E}_{\alpha}(\mathbf{r})\ \mathcal{E}^\dagger_\beta (\mathbf{r}')
\label{appeq: gf_spectral_form_wg_cav}
\end{align}
where for propagating modes
\begin{equation}
\mathcal{H} \ket{\mathcal{E}_k} =\frac{\omega_k}{c}\underbrace{(\bar{\varepsilon}_b(\mathbf r)+\Delta\bar{\varepsilon}_w
(\mathbf{r}))}_{\bar{\varepsilon}_k} \ket{\mathcal{E}_k}
\end{equation}
and for the stationary modes
\begin{equation}
\mathcal{H} \ket{\mathcal{E}_c} =\frac{\omega_c}{c}\underbrace{(\bar{\varepsilon}_b(\mathbf r)+\Delta\bar{\varepsilon}_c
(\mathbf{r}))}_{\bar{\varepsilon}_c} \ket{\mathcal{E}_c}
\end{equation}

Then projecting the eigen equations onto the modes, give the interaction strength between the $k^{\mathrm{th}}$ propagating mode and $c^{\mathrm{th}}$ stationary mode and vice-versa, 
\begin{equation}
    \begin{split}
    J_{k,c} = c\langle\mathcal E_k|\mathcal H|\mathcal E_c\rangle  = \omega_c \langle\mathcal E_k|\bar{\varepsilon}_c|\mathcal E_c\rangle\\
    J_{c,k} = c\langle\mathcal E_c|\mathcal H|\mathcal E_k\rangle  = \omega_k \langle\mathcal E_c|\bar{\varepsilon}_k|\mathcal E_k\rangle
    \end{split}
\end{equation}
where for overlapping modes of interest $\omega_c\approx\omega_k$ will give $J_{k,c} = J_{c,k}^\dagger$. By invoking the coupled GF relation and the modal expansions, one can deduce the coupled relations
\begin{equation}
A_{k,c}(\omega) (\frac{\omega_k-\omega}{c}) - A_{c,c}(\omega)\frac{\omega}{c} \sum_{k} J_{k,c} = 0
\label{appeq: kc_cc_relation_1}
\end{equation}

\begin{equation}
\frac{\omega_c-\omega}{c} A_{c,c}(\omega) - \frac{\omega}{c}A_{k,c}(\omega) \sum_{k} J_{c, k} = \frac{\omega}{c}
\label{appeq: kc_cc_relation_2}
\end{equation}
Solving for each source index gives all four types of coefficient:
\begin{equation}
\begin{aligned}
A_{cc}&=-\frac{c}{D_c},&
A_{kc}&=-\frac{cJ_{k,c}}{\Delta_k D_c},\\
A_{ck}&=-\frac{cJ_{k,c}^*}{D_c \Delta_k},&
A_{kk'}&=-\frac{c\delta_{kk'}}{\Delta_k}
-\frac{cJ_{k,c}J_{k',c}^*}{\Delta_k D_c \Delta_{k'}}.
\end{aligned}
\label{eq:tmpcomp_all_coefficients}
\end{equation}
where $\Delta_{k(c)} =\omega - \omega_{k(c)}$ and $D_c = \Delta_{c} -\Sigma_{\mathrm wg}(\omega)$ with $\Sigma_{\mathrm wg}(\omega)$ representing the modification in cavity's self-energy due to its dressing with the propagating modes of the waveguide, defined as, 
\begin{equation}
\Sigma_{\rm wg}(\omega)=\sum_k\frac{|J_{k,c}|^2}{\Delta_k}
\label{eq:tmpcomp_self_energy}
\end{equation}

Therefore, equation~\eqref{appeq: gf_spectral_form_wg_cav} can be decomposed neatly into, 
\begin{equation}
g\simeq g_{\rm wg}+g_{\rm cav}^{\rm dressed}+g_{\rm coupled},
\label{eq:tmpcomp_green_parts}
\end{equation}
where
\begin{equation}
\begin{aligned}
g_{\rm wg}&=-c\sum_k\frac{|\mathcal E_k\rangle\langle\mathcal E_k|}{\Delta_k},\\
g_{\rm cav}^{\rm dressed}&=-\frac{c}{D_c}
|\mathcal E_c\rangle\langle\mathcal E_c|,
\end{aligned}
\end{equation}
and the coupling contribution is
\begin{equation}
\begin{aligned}
g_{\rm coupled}=-\frac{c}{D_c}\Bigg[&
\sum_k\frac{J_{k,c}|\mathcal E_k\rangle\langle\mathcal E_c|
+J_{c,k}|\mathcal E_c\rangle\langle\mathcal E_k|}{\Delta_k}\\
&+\sum_{k,k'}\frac{J_{k,c}J_{c,k'}
|\mathcal E_k\rangle\langle\mathcal E_{k'}|}{\Delta_k\Delta_{k'}}\Bigg].
\end{aligned}
\label{eq:tmpcomp_coupled_green}
\end{equation}
The first sum describes propagation from cavity to waveguide and from waveguide to cavity. The double sum describes propagation into the cavity from one waveguide mode and back into another. It is present even without direct waveguide-mode mixing and cannot generally be omitted near the dressed cavity resonance. All denominators are retarded; conjugating a coupling coefficient does not conjugate its frequency denominator.

Finally, take $\sum_k\to L\int dk/(2\pi)$ for box length $L$. The continuum self-energy has the form
\begin{equation}
\begin{aligned}
\Sigma_{\rm wg}(\omega)&=\Delta_{\rm wg}(\omega)
-i\kappa_{\rm wg}(\omega)/2,\\
\kappa_{\rm wg}(\omega)&=2\pi\sum_k|J_{k,c}|^2\delta(\omega-\omega_k),\\
\Delta_{\rm wg}(\omega)&=\mathcal P\sum_k
\frac{|J_{k,c}|^2}{\omega-\omega_k}.
\end{aligned}
\label{eq:tmpcomp_shift_linewidth}
\end{equation}
The delta function supplies a factor $1/|v_g|$ at each propagating root; the box-normalized $J_{k,c}$ scales as $L^{-1/2}$, leaving a finite limit. Consequently $D_c=\omega-\omega_c-\Delta_{\rm wg}+i(\kappa_{\rm int}+\kappa_{\rm wg})/2$. A slowly varying self-energy gives the familiar cavity frequency shift and additional waveguide linewidth; otherwise its full frequency dependence is retained. These expressions are exact within the stated reduced model. To obtain modal scattering, insert the resulting Green function into the same boundary projection used in the main text, retaining any background outside this modal subspace.

\section{Numerical Green-function reconstruction}
\label{app:numerical_reconstruction}

We performed FDTD simulations of the dielectric structures using Tidy3D~\cite{tidy3d}. A point electric-current source with moment amplitude $d_{\rm sim}=1\,\mathrm{A\,\mu m}$ was placed at the emitter position $\mathbf r_A$ and polarized along the transition-dipole direction $\hat{\mathbf e}$. Here $d_{\rm sim}$ is a current dipole moment, distinct from the emitter's electric transition dipole $\mathbf d$. The source-normalized electric field determines the corresponding column of the second-order Green tensor through
\begin{equation}
g^{(2)}(\mathbf r,\mathbf r_A,\omega)\hat{\mathbf e}
=\frac{\mathbf E(\mathbf r,\omega)}{i\mu_0\omega d_{\rm sim}}.
\label{eq:num_sim_source_conversion}
\end{equation}
We recorded the field at the source position to obtain the self-response and at reference points in the input and output waveguides to obtain the propagation amplitudes.

For the block-diagonal material response of Eq.~\eqref{eq:dual_block_operators}, the first-order Green tensor can be written in terms of the electric and magnetic second-order Green tensors~\cite{partI}. With curls acting on the observation coordinate $\mathbf r$, the relation is
\begin{equation}
g=
\begin{pmatrix}
k_0g_E^{(2)} & i\boldsymbol\varepsilon^{-1}(\mathbf r)\nabla_{\mathbf r}\times g_H^{(2)}\\[3pt]
-i\boldsymbol\mu^{-1}(\mathbf r)\nabla_{\mathbf r}\times g_E^{(2)} & k_0g_H^{(2)}
\end{pmatrix}.
\label{eq:num_full_green_conversion}
\end{equation}
All Green tensors have arguments $(\mathbf r,\mathbf r',\omega)$ and the same retarded boundary prescription; $\boldsymbol\varepsilon$ and $\boldsymbol\mu$ are the relative material tensors. Here $g_E^{(2)}$ and $g_H^{(2)}$ solve the electric and magnetic vector Helmholtz equations, respectively. The tensor denoted by $g^{(2)}$ in the main text is $g_E^{(2)}$. Electric-dipole self-energies select only the upper-left block, $g_{EE}=k_0g_E^{(2)}$.

To obtain the emitter self-response, we evaluated the electric field at the dipole position and repeated the calculation in vacuum using the same mesh and source placement. Subtracting the vacuum Green tensor gives the structure-induced response, $g_{\rm sc}^{(2)}=g^{(2)}-g_0^{(2)}$~\cite{van2012DyadGF}. When the local medium at the source is unchanged, this subtraction removes the common singular self-field. Because the real self-response is sensitive to spatial discretization, we do not interpret it as an absolute dielectric Lamb shift. For the cavity and atom-coupler examples, we fit the complex, dipole-projected scattered response to a resonant pole plus a smooth linear background and retain the pole contribution to the plotted frequency shift. The decay rate is calculated from the full imaginary self-response, without subtracting the vacuum contribution.

The simulation source amplitude $d_{\rm sim}$ fixes the Green-function normalization but does not set the emitter's transition dipole. We choose the latter through the free-space population-decay rate $\Gamma_0$. For a transition polarized along $\hat{\mathbf e}$, define the projected vacuum response $g_0^{(2)}(\omega)=\hat{\mathbf e}^{\dagger}g_{\rm vac}^{(2)}(\mathbf r_A,\mathbf r_A,\omega)\hat{\mathbf e}$. The required dipole strength is then
\begin{equation}
|d|^2=\frac{\hbar\Gamma_0}
{2\mu_0\omega_A^2\operatorname{Im}g_0^{(2)}(\omega_A)}.
\label{eq:num_transition_dipole_calibration}
\end{equation}
We use the simulated vacuum response for this calibration and choose $\Gamma_0/(2\pi)=30\,\mathrm{GHz}$. This free-space linewidth is held fixed for each emitter detuning and for all three structures.

With the transition dipole fixed by $\Gamma_0$, the normalized total decay rate and resonant frequency shift are
\begin{equation}
\begin{aligned}
\frac{\Gamma_A(\omega)}{\Gamma_0}
&=\left(\frac{\omega}{\omega_A}\right)^2
\frac{\operatorname{Im}g_{\rm self}^{(2)}(\omega)}
{\operatorname{Im}g_0^{(2)}(\omega_A)},\\
\frac{\Delta_{\rm res}(\omega)}{\Gamma_0}
&=-\frac12\left(\frac{\omega}{\omega_A}\right)^2
\frac{\operatorname{Re}g_{\rm res}^{(2)}(\omega)}
{\operatorname{Im}g_0^{(2)}(\omega_A)}.
\end{aligned}
\label{eq:num_normalized_decay_shift}
\end{equation}
Here all Green functions are projected along the transition-dipole direction, and $g_{\rm res}^{(2)}$ denotes the fitted pole contribution to the scattered self-response. The decay includes all electromagnetic channels, whereas the shift retains only the resonant contribution. We use this same resonant-shift approximation when calculating transmission for the cavity and atom-coupler examples.

To display the emitter-mediated coupling relative to the bare transmission phase, we plot the dimensionless quantity
\begin{equation}
Q(\omega)=\frac{\eta_{\rm out}(\omega)\eta_{\rm in}(\omega)}
{\Gamma_A(\omega)}e^{-i\phi_t(\omega)},
\qquad \phi_t(\omega)=\arg t(\omega).
\label{eq:num_phase_coupling}
\end{equation}
This removes the phase of the bare transmission, but not its magnitude. The magnitude of $Q$ measures the channel-coupling product relative to the total emitter decay rate, while its phase determines the interference with the bare transmission after including the emitter-response denominator. The phase reference is defined where $t(\omega)\ne0$.

At the shifted emitter resonance, the relative transmission is governed by the dimensionless ratio $g_{\rm out}^{(2)}g_{\rm in}^{(2)}/[g_{\rm in-out}^{(2)}\operatorname{Im}g_{\rm self}^{(2)}]$, where the Green functions are consistently normalized, polarization-projected propagation amplitudes. This compares the emitter-mediated and direct propagation paths relative to the total electromagnetic decay; its magnitude and phase determine the resulting extinction or enhancement. The explicit transition-dipole strength cancels from this ratio, provided the reference points admit single-mode reconstruction and there is no additional emitter broadening.

\bibliography{references}
\bibliographystyle{apsrev4-2}

\end{document}